\documentclass[12pt]{iopart}
\usepackage{amscd,amssymb}

\usepackage{epsfig}
\usepackage{amscd,amssymb}
\usepackage{graphicx,xspace}
\usepackage{pstricks}
\usepackage[left=3cm,top=3cm,right=3cm,bottom=3cm]{geometry}
\usepackage{color}
\definecolor{labelkey}{cmyk}{.4,.2,0,0}

\begin{document}

\newcommand \be  {\begin{equation}}
\newcommand \bea {\begin{eqnarray} \nonumber }
\newcommand \ee  {\end{equation}}
\newcommand \eea {\end{eqnarray}}

\title{A spin glass model for reconstructing nonlinearly encrypted signals corrupted by noise.}

\vskip 0.2cm

\author{Yan V Fyodorov}

\address{ King's College London, Department of Mathematics, London  WC2R 2LS, United Kingdom}

\begin{abstract}
We define a (symmetric key) encryption of a signal ${\bf s}\in\mathbb{R^N}$ as a random mapping ${\bf s}\mapsto \textbf{y}=(y_1,\ldots,y_M)^T\in \mathbb{R}^M$ known both to the sender and a recipient. In general the recipients may have access only to images ${\bf y}$ corrupted by an additive noise. Given the Encryption Redundancy Parameter (ERP) $\mu=M/N\ge 1$, the signal strength parameter $R=\sqrt{\sum_i s_i^2/N}$, and the ('bare') noise-to-signal ratio (NSR) $\gamma\ge 0$, we consider the problem of reconstructing ${\bf s}$ from its corrupted image  by a Least Square Scheme for a certain class of random Gaussian mappings. The problem is equivalent to finding the configuration of minimal energy in a certain version of spherical spin glass model, with {\it squared} Gaussian random interaction potential. We use the Parisi replica symmetry breaking scheme to evaluate the mean overlap $p_{\infty}\in [0,1]$ between the original signal and its recovered image (known as 'estimator') as $N\to \infty$, which is a measure of the quality of the signal reconstruction. We explicitly analyze the general case of {\it linear-quadratic} family of random mappings and discuss the full $p_{\infty} (\gamma)$ curve. When nonlinearity exceeds a certain threshold but redundancy is not yet too big, the replica symmetric solution is necessarily broken in some interval of NSR. We show that encryptions with a nonvanishing linear component permit reconstructions with $p_{\infty}>0$ for any $\mu>1$ and any $\gamma<\infty$, with $p_{\infty}\sim \gamma^{-1/2}$ as $\gamma\to \infty$. In contrast, for the case of {\it purely quadratic} nonlinearity, for any ERP $\mu>1$ there exists a threshold NSR value $\gamma_c(\mu)$ such that $p_{\infty}=0$ for $\gamma>\gamma_c(\mu)$ making the reconstruction impossible.
   The behaviour close to the threshold is given by $p_{\infty}\sim (\gamma_c-\gamma)^{3/4}$ and is controlled by the replica symmetry breaking mechanism.

\end{abstract}

\maketitle

\section{Introduction}

 In this paper we consider a schematic model of a (symmetric key) reconstruction of a source  signal from its encrypted form corrupted by an additive noise when passed from a sender to a  recipient. Signals are represented by  $N-$dimensional source (column)  vectors $\textbf{s}={\small \left(\begin{array}{c}s_1 \\\ldots \\ s_N\end{array}\right)}\in \mathbb{R}^N$, and we define the associated   signal strength $R$  via the Euclidean norm as $R=\sqrt{\frac{1}{N}\left({\bf s},{\bf s}\right)}$, where $(\cdot,\cdot)$ stands for the Euclidean inner product in $\mathbb{R}^N$.
By a (symmetric key) {\it encryption} of the source signal we understand a {\it random mapping} ${\bf s}\mapsto \textbf{y}={\small \left(\begin{array}{c}y_1 \\\ldots \\ y_M\end{array}\right)}\in \mathbb{R}^M$ known both to the sender and a recipient.  For further reference we find it useful to write the mapping component-wise explicitly as
\begin{equation}
y_k=V_k(\textbf{s}), \quad k=1,\ldots, M\,,
\end{equation}
with the collection of random functions $V_1({\bf s}), \ldots, V_M(\textbf{s})$ representing an encryption algorithm shared between the parties participating in the signal exchange. Due to imperfect communication channels the recipients however get access to the encrypted signals only in a corrupted form ${\bf z}$. We consider only the simplest corruption mechanism when the encrypted images  ${\bf y}$ are modified by an additive random noise, i.e.  $\textbf{z}=\textbf{y}+{\bf b}$.  The noise vectors ${\bf b}$ are further assumed to be
normally distributed: ${\bf b}\sim {\cal N}({\bf 0},\sigma^2 {\bf 1}_M)$, i.e. components $b_k,\,k=1,\ldots,M$  are i.i.d. mean zero real Gaussian variables with the covariance $\left\langle b_kb_l \right\rangle=\delta_{kl}\sigma^2$, where the notation $\left\langle \ldots \right \rangle $ here and henceforth stands for the expected value $\mathbb{E}[\ldots]$ with respect to all types of random variables.  A natural parameter  is then the 'bare' noise-to-signal ratio (NSR) $\gamma=\sigma^2/R^2$, which will be eventually converted to true NSR dependent on the parameters of encryption algorithm
(we will later on refer to such conversion in the text as an appropriate 'scaling') characterizing the level of signal corruption
in the chosen type of encryption.

The recipient's aim is to reconstruct the source signal ${\bf s}$ from the knowledge of ${\bf z}$. In the presence of noise
such reconstruction can be only approximate, and reconstructed signals are known in the signal processing literature as 'estimators' of the source signals. Their properties depend on the reconstruction scheme used.
 In the Bayesian inference approach philosophy one exploits reconstruction schemes optimized, among other parameters, over the probabilities of the {\it input signal} ${\bf s}$  by postulating its {\it prior} distribution over the set of feasible input signals. In that way one of the most popular estimators is the minimum mean square error (MMSE) estimator.
  We do not follow the Bayesian approach here, and rather consider the input signal through the reconstruction procedure as a  {\it fixed} vector, and then employ the Least-Square reconstruction scheme, which returns an estimator as
\begin{equation}\label{leastsquare}
{\bf x}:=Argmin_{{\bf w}}\,\left[\sum_{k=1}^M \frac{ \left(z_k-V_k(\bf{w})\right)^2}{2}\right], \quad {\bf w}\in \mathbb{W}\subseteq  \mathbb{R}^N
\end{equation}
where $\mathbb{W}$ is a set of feasible signals. Since for a given input ${\bf s}$ and the Gaussian noise ${\bf b}$ the probability $p({\bf z}|{\bf s})$ to observe ${\bf z}$ is given by $p({\bf z}|{\bf s})\sim \exp\left\{-\frac{1}{\sigma^2}\sum_{k=1}^M \frac{ \left(z_k-V_k(\bf{s})\right)^2}{2}\right\}$, the estimator Eq.(\ref{leastsquare}) is also known as the {\it maximum likelihood}
estimator. Note however that this approach can be given a formal Bayesian meaning as a Maximum--A-Posteriori
(MAP) estimator with a uniform prior distribution over the feasibility set, see below.
 Quality of the signal reconstruction under this scheme  is then characterized by the value of a distortion parameter measuring the difference between the fixed source signal ${\bf s}$ and the reconstructed estimator ${\bf x}$. For this one can use any suitable distance function
$d({\bf x};{\bf s})$, e.g the Euclidean distance normalized to the signal strength:
\begin{equation}\label{dist}
d({\bf x};{\bf s})=\frac{\left({\bf x}-{\bf s}\right)^2}{NR^2}
\end{equation}
One is interested in getting the expression for the distortion in the asymptotic limit of large signal dimensions $(M,N)\to \infty$.
As long as $M$ remains smaller than $N$, any solution ${\bf x}\in \mathbb{W}\subseteq  \mathbb{R}^N$ of the set of $M$ equations $b_k+V_k({\bf s})-V_k({\bf x})=0,\,\, k=1,\ldots, M$ will be corresponding to the eactly zero value of the cost function, and
could be a legitimate estimator. Those estimators then form continuously parametrized manifolds in $\mathbb{R}^N$.
It is therefore clear that even in the absence of any noise full reconstruction of the signal for $M<N$  under this scheme is impossible.
Although such a case is not at all devoid of interest, we do not treat it in the present paper leaving it for a separate study.   In contrast, for 'redundantantly' encrypted signals with  $M\ge N$ the set of possible estimators generically consists of isolated points in $\mathbb{R}^N$. To this end we introduce the  Encryption Redundancy Parameter (ERP) $\mu=M/N\in(1,\infty)$. We will see that under such conditions
signals can be in general faithfully  reconstructed in some range of the  noise-to-signal ratios $\gamma>0$.

In this paper we are going to apply  tools of Statistical Mechanics for calculating the average asymptotic distortion for a certain class of the least square reconstruction of a  randomly encrypted noisy signal. As this is essentially a large-scale random optimization problem, methods of statistical mechanics of disordered systems like the replica trick developed in theory of spin glasses are known to be efficient in providing important analytical insights in the statistical properties of the solution, see e.g. \cite{MPV,Parisilec04}.  It is also worth noting that distinctly different aspects of the problem of information reconstruction ( the so-called {\it error-correcting} procedures) were discussed in the framework of spin glass ideas already in the seminal work by Sourlas \cite{Sourlas}.

We consider the reconstruction problem  under two technical assumptions.
The first assumption is that the recipient is aware of the exact source signal strength $R$, and therefore can restrict the least square minimization search in Eq.(\ref{leastsquare}) to the feasibility set $\mathbb{W}$ given by  $(N-1)-$dimensional sphere of the radius $R\sqrt{N}$. We will refer to such a condition as the 'spherical constraint'. From the point of view of the Bayesian
analysis our reconstruction scheme can be considered as a MAP estimator with postulated prior distribution
being the uniform measure on the above-mentioned $(N-1)-$dimensional sphere. As the lengths of both the input signal ${\bf s}$
and an estimator ${\bf x}$ are fixed to $R\sqrt{N}$, the distance (\ref{dist}) depends only on the scalar product
$({\bf x},{\bf s})$. We therefore can conveniently characterize the quality of the reconstruction via the {\it quality parameter} defined as
\begin{equation}\label{qualirec}
p_N=\frac{({\bf x},{\bf s})}{NR^2}\in [0,1]
\end{equation}
where $p_N=1$ corresponds to a reconstruction without any macroscopic distortion, whereas $p_N=0$ manifests impossibility to recover any information from the originally encrypted signal. Note that the assumption of the fixed input signal strength $R$ is technically convenient but can be further relaxed; the analysis can be extended, without much difficulty, to the search in a spherical shell $R_1\sqrt{N}\le |{\bf w}| \le R_2\sqrt{N}$, and hopefully to some other situations.

 Our second assumption is that the random functions $V_k({\bf s})$ belong to the class of (smooth) {\it isotropic} gaussian-distributed random fields on the sphere, independent for different values of $k$ (and independent of the noise ${\bf b}$)
, with mean-zero and the covariance structure dependent only on the angle between the vectors. Using the scaling appropriate for our problem in the limit of large $N$ we represent such covariances as
\begin{equation}\label{cov}
\left\langle V_k({\bf x})V_l({\bf s})\right\rangle=\delta_{lk}\Phi\left(\frac{({\bf x},{\bf s})}{N}\right)
\end{equation}
where the angular brackets $\left\langle \ldots \right\rangle$  denote the expectation
with respect to the corresponding probability measures. We will further assume for simplicity that $\Phi\left(u\right)$ in
Eq.(\ref{cov}) is infinitely differentiable.

The simplest case of random fields of this type corresponds to a {\it linear} encryption algorithm, with the functions $V_k(\textbf{x}), \quad k=1,\ldots, M$  chosen in the form of random linear combinations:
\begin{equation}\label{linear}
V_k(\textbf{x})=\sum_{i=1}^N a_{ki}x_i:=({\bf a}_k,{\bf x}),
\end{equation}
where the vectors ${\bf a}_k$ are assumed to be random, mean-zero mutually independent Gaussian, each with $N$ i.i.d. components characterized by the variances $\left\langle a_{ki}a_{lj}\right\rangle=\frac{J_1^2}{N}\delta_{lk}\delta_{ij}$. Such choice
 implies the covariance (\ref{cov}) with $\Phi\left(u\right)=J_1^2\, u$.  

The linear encryption is very special, yet not completely trivial, instance of the reconstruction problem,  as in that case one can formally solve the minimization problem by the method of Lagrange multipliers explicitly. To this end we introduce the cost function (cf. (\ref{leastsquare}))
\begin{equation}\label{Hamilt}
 {\cal H}_{\bf s}({\bf x})=\sum_{k=1}^M   {\cal H}^{(k)}_{\bf s}({\bf x}), \quad {\cal H}^{(k)}_{\bf s}({\bf x})=\frac{\left({\bf b}+V_k({\bf s})-V_k(\bf{x})\right)^2}{2}
\end{equation}
depending on the source signal ${\bf s}$ as a parameter, and following the standard idea of a constrained minimization consider the stationarity conditions $\nabla {\cal L}_{\lambda,{\bf s}}({\bf x})=0$ for the Lagrangian ${\cal L}_{\lambda,{\bf s}}({\bf x})={\cal H}_{\bf s}({\bf x})-\frac{\lambda}{2}({\bf x},{\bf x})$, with real $\lambda$ being the Lagrange multiplier taking care of the spherical constraint.  In the general case of a non-linear encryption algorithm this procedure does not seem to help much to our analysis, as the  stationarity equations look hard to study. In the linear case one can however introduce a $N\times M$ matrix $A$ whose $M$ rows are represented by (transposed) vectors ${\bf a}_k^T$ featuring in Eq.(\ref{linear}). We than can easily see that the stationarity conditions in that case amount to the following matrix equation:
\begin{equation}\label{Lagrange}
 A^T\left[A({\bf x}-\bf{s})-{\bf b}\right]=\lambda {\bf x}
\end{equation}
which can be then immediately solved and provides the estimator in the form
\begin{equation}\label{Lagrangesol}
{\bf x}=\left[{\bf 1}_N+\lambda \left(A^TA-\lambda {\bf 1}_N\right)^{-1} \right]{\bf s}+\left(A^TA-\lambda {\bf 1}_N\right)^{-1}A^T{\bf b}
\end{equation}
The possible set of Lagrange multipliers is obtained by solving the equation implied by the spherical constraint:
$({\bf x},{\bf x})=NR^2$, which is in general equivalent to a polynomial equation of degree $2N$ in $\lambda$.
The number of real solutions of that equation depends on the noise vector ${\bf b}$. One of the real solutions corresponds to the
minimum of the cost function, others to saddle-points or maxima. In particular, in the (trivial) limiting case of no noise ${\bf b}=0$
the global minimum corresponds to $\lambda=0$ implying reconstruction with no distortion:  ${\bf x}={\bf s}$, hence $p_N=1$ as is natural to expect.
At the same time, for any ${\bf b}\ne 0$ the analysis of Eq.(\ref{Lagrangesol}) becomes a non-trivial problem. One possible way is to
account for the presence of a weak noise with small variance $\sigma^2$ by developing a perturbation theory in the small scaled NSR parameter $\tilde{\gamma}=\frac{\sigma^2}{J_1^2R^2}\ll 1$. Such a theory is outlined in the Appendix A, where we find that for a given value of ERP $\mu>1$ and in the leading order in $\tilde{\gamma}$ the asymptotic disorder-averaged quality reconstruction parameter defined in Eq.(\ref{qualirec}) is given by:
\begin{equation}\label{qualiasylin}
p_{\infty}:=\lim_{N\to \infty}\left\langle p_N\right\rangle=1-\tilde{\gamma}\frac{1}{\mu-1} +o(\tilde{\gamma})
\end{equation}
This result is based on the asymptotic mean density of eigenvalues of random Wishart matrices $A^TA$  due to Marchenko and Pastur \cite{MP}.  Similarly, one can develop a perturbation theory for very big NSR $\tilde{\gamma}\gg 1$, see Appendix A. In this way one finds that the Lagrange multiplier $|\lambda|\sim (\mu\tilde{\gamma})^{1/2}$ and $p_{\infty}\sim \sqrt{\frac{\mu}{\tilde{\gamma}}}$.

Although the perturbation theories are conceptually straightforward, and can be with due effort extended to higher orders, the calculations quickly become too cumbersome. At the moment we are not aware of any direct approach to our minimization problem in the linear encryption case which may provide non-perturbative results, as $N\to \infty$, for asymptotic distortion $p_{\infty}$ at values of scaled NSR parameter $\tilde{\gamma}$ of the order of unity. At the same time we will see that methods of statistical mechanics provide a very explicit expression for any $\tilde{\gamma}$.

 It is necessary to mention that various instances of not dissimilar {\it linear} reconstruction problems in related forms received recently a considerable attention.  The emphasis in those studies seems however to be mainly restricted to the case  of source signals being subject to a compressed sensing, i.e. represented by a sparse vector with a finite fraction of zero entries, see e.g. \cite{Krzakala2012,ZKrev,Montanari16} and references therein. To this end especially deserve mentioning the works \cite{BMSB1,BMSB2,BerMue} which studied the mean value of distortions for MAP estimator for a linear problem (though with prior distribution different from the spherical constraint). Although having a moderate overlap with methods used in this work, the actual calculations and the main message of those papers seem rather different.

 In particular, our main emphasis will be on ability to analyse the case of a quite general {\it nonlinear} random Gaussian encryptions\footnote{
 In the context of compressed sensing some reconstruction aspects of nonlinear models were considered, e.g. in \cite{nonlinsensing1,nonlinsensing2}, but our approach seems distinctly different.}.
The corresponding class of functions $V_k(\textbf{x})$ extends the above-mentioned case of random linear forms  to higher-order random forms, the first nontrivial example being the form of degree 2:
\begin{equation}\label{linearquadratic}
V_k(\textbf{x})=({\bf a}_k,{\bf x})+\frac{1}{2}({\bf x},{\cal J}^{(k)}{\bf x})
\end{equation}
where entries of $N\times N$ real symmetric  random matrices ${\cal J}^{(k)}, k=1,\ldots, M$  are mean-zero Gaussian variables (independent of the vectors ${\bf a}_k$) with the covariance structure
\begin{equation}\label{quadcov}
\left\langle {\cal J}^{(k)}_{ij} {\cal J}^{(l)}_{mn}\right\rangle=\frac{J_2^2}{N^2}\delta_{lk}\left(\delta_{im}\delta_{jn}+\delta_{in}\delta_{jm}\right)
\end{equation}
which eventually results in the covariance (\ref{cov}) of the form $\Phi\left(u\right)=J_1^2\, u+\frac{1}{2}J_{2}^2u^2$.
We will refer to the above class of random encryptions as the {\it linear-quadratic family}.

  In fact, the general covariance structure of isotropic Gaussian random fields on a sphere of radius $R\sqrt{N}$ is also well-known from the theory of spherical spin glasses: these are functions which can be represented by a (possibly, terminating) series with non-negative coefficients:
 \begin{equation}\label{gen-nonlin}
 \Phi\left(u\right)=\sum_{l=1}^{\infty} c_l u^{l}, \quad c_l\ge 0\, \, \forall l
 \end{equation}
  such that $\Phi\left(R^2\right)$ has a finite value, see e.g. \cite{Auf2}. Although our theory of encrypted signal
reconstruction will be developed for the general case, all the explicit analysis of the ensuing equations will be restricted to the case of the linear-quadratic family, Eq.(\ref{linearquadratic}).

\subsection{\sf Main Results}
 Our first main result is the following\\
 {\bf Proposition 1}: {\it Given a value of $R>0$ characterizing the source signal strength, and the value of the Encryption Redundancy Parameter $\mu>1$,  consider the functional
\begin{equation}\label{mainzerotemp}
\fl \mathcal{E}[w_s(u);Q,v,t]=
-\left[\frac{R^2-t^2-Q}{v+\int_{R^2-Q}^{R^2}w_s(u)\,du}+\int_{R^2-Q}^{R^2}\frac{dq}{v+\int_{q}^{R^2}w_s(u)\,du}\right]
\end{equation}
\[
\fl +\mu\left[\frac{\sigma^2+\Phi(R^2)-2\Phi(Rt)+\Phi(R^2-Q)}{1+v\Phi'(R^2)+\int_{R^2-Q}^{R^2}w_s(u)\Phi'(u)\,du }
+\int_{R^2-Q}^{R^2}\frac{\Phi'(q)\,dq}{1+v\Phi'(R^2)+\int_{q}^{R^2}w_s(u)\Phi'(u)\,du }\right]
\]
where the variable $v\ge 0$, the variables $t$ and $Q$  take values in intervals $[-R,R]$ and $[0,R^2]$, correspondingly,
and  $w_s(u)$ is a non-decreasing function in $u\in[R^2-Q,R^2]$.
      Then in the framework of the Parisi scheme of the replica trick the mean value of the parameter $p_N$ characterising quality of the information recovery in the signal reconstruction scheme, Eqs.(\ref{leastsquare})-(\ref{cov}), with normally distributed noise ${\bf b}\sim {\cal N}({\bf 0},\sigma^2 {\bf 1}_M)$  is given for $N\to \infty$ by
    \be\label{qualiasy}
p_{\infty}:=\lim_{N\to \infty}\left\langle p_{N}\right\rangle = \frac{t}{R}
\ee
    where the specific value of the parameter $t$ to be substituted to (\ref{qualiasy}) should be found by simultaneously minimizing the functional $\mathcal{E}[w_s(u);Q,v,t]$ over $t$ and
 maximizing it over all other parameters and the function $w_s(u)$.
 }

\vspace{0.5cm}

Our next result provides an explicit solution to this variational problem in a certain range
of parameters.\\

{\bf Proposition 2}: {\it

In the range of parameters such that the solution $t$ to the equation
\begin{equation}\label{stat-Q1-fin-AT}
 \mu\,R^2\left[\Phi'(Rt)\right]^2(R^2-t^2)=t^2\Phi'(R^2)\left[R^2\gamma+\Phi(R^2)-2\Phi(Rt)+\Phi(R^2)\right]
\end{equation}
satisfies the inequality
\begin{equation}\label{AT-line-gen}
\left[\Phi'(R^2)\right]^3\,\frac{t^2}{R^2} \le  \mu \left[\Phi'(Rt)\right]^2 \left[\Phi'(R^2)-\Phi''(R^2)(R^2-t^2)\right]
\end{equation}
the variational problem Eq.(\ref{mainzerotemp}) is solved by the Replica-Symmetric Ansatz $Q=0$. In particular,
for a given 'bare' Noise-to-Signal ratio $\gamma=\sigma^2/R^2$ the quality parameter $p_{\infty}$ definied in Eq.(\ref{qualiasy})
is given by the solution of the following equation:
\begin{equation}\label{p-RS-Main}
p^2\left(\gamma+2\,\frac{\Phi(R^2)-\Phi(R^2p)}{R^2}\right)=\mu (1-p^2)\frac{\left[\Phi'(R^2p)\right]^2}{\Phi'(R^2)}
\end{equation}
In addition, for the range of parameters such that the solution $t$ of Eq.(\ref{stat-Q1-fin-AT}) violates the inequality
Eq.(\ref{AT-line-gen}) the variational problem Eq.(\ref{mainzerotemp}) is solved by the Full Replica-Symmetry Breaking Ansatz.
In this case the value $p$ of the quality parameter $p_{\infty}$ definied in Eq.(\ref{qualiasy})
is given by the solution of the following system of two equations in the variables $p\in[0,1]$ and $Q\in [0,R^2]$:
\begin{equation}\label{RSB-Main-1}
\fl \mu\,\left[\Phi'(R^2p)\right]^2\left(R^2(1-p^2)-Q)\right)=p^2\Phi'(R^2-Q)\left[R^2\gamma+\Phi(R^2)-2\Phi(R^2p)+\Phi(R^2-Q)\right]
\end{equation}
\begin{equation}\label{RSB-Main-2}
\fl \left[\Phi'(R^2-Q)\right]^3\,p^2=\mu \left[\Phi'(R^2p)\right]^2 \left[\Phi'(R^2-Q)-\Phi''(R^2-Q)\left(R^2(1-p^2)-Q\right)\right]
\end{equation}
}
We finally note in passing that  an attempt to extremize the functional Eq.(\ref{mainzerotemp}) in the space of the so-called 1-Step Replica Symmetry Breaking Ansatz (1-RSB) does not yield any solution respecting the required constraints on the parameters $v,t$ and $Q$.

\subsubsection{Results for the linear-quadratic family of encryptions.}

Both Propositions providing the solution of our reconstruction problem in full generality, for every specific choice of the covariance structure $\Phi(u)$ the equations need to be further analyzed.
In this work we performed a detailed analysis of the case of encryptions belonging to the linear-quadratic family Eq.(\ref{linearquadratic}) with the covariance structure of the form $\Phi\left(u\right)=J_1^2\, u+\frac{1}{2}J_{2}^2u^2$.
The most essential  qualitative features of the analysis are summarized below.

For such a family, apart from our main control parameters, scaled NSR $\tilde{\gamma}=\gamma/J_1^2$ and ERP $\mu$, the reconstruction  is very essentially controlled
by  an important parameter $a=R^2J_{2}^2/J_1^2$ which reflects the degree of {\it nonlinearity} in the encryption mapping.
Our first result is that there exists a threshold value of this parameter, $a=1$, such that for all encryptions in the family
with $a<1$ the variational problem is always solved with the Replica-Symmetric Ansatz Eq.(\ref{p-RS-Main}).
In contrast,  for linear-quadratic encryptions with higher nonlinearity $a>1$ there exists a threshold value of the Encryption Redundancy Parameter $\mu=\frac{(a^{2/3}-a^{1/3}+1)^3}{a}:=\mu_{AT}(a)>1$ such that for any $\mu\in(1,\mu_{AT}(a))$
 the replica symmetric solution is broken in some interval of NSR. This implies that increasing redundancy for a fixed non-linearity
 one eventually always ends up in the replica-symmetric phase, see the phase diagram in Fig.1.

 \begin{figure}[h!]
 \centering
  \includegraphics[width=0.6\textwidth]{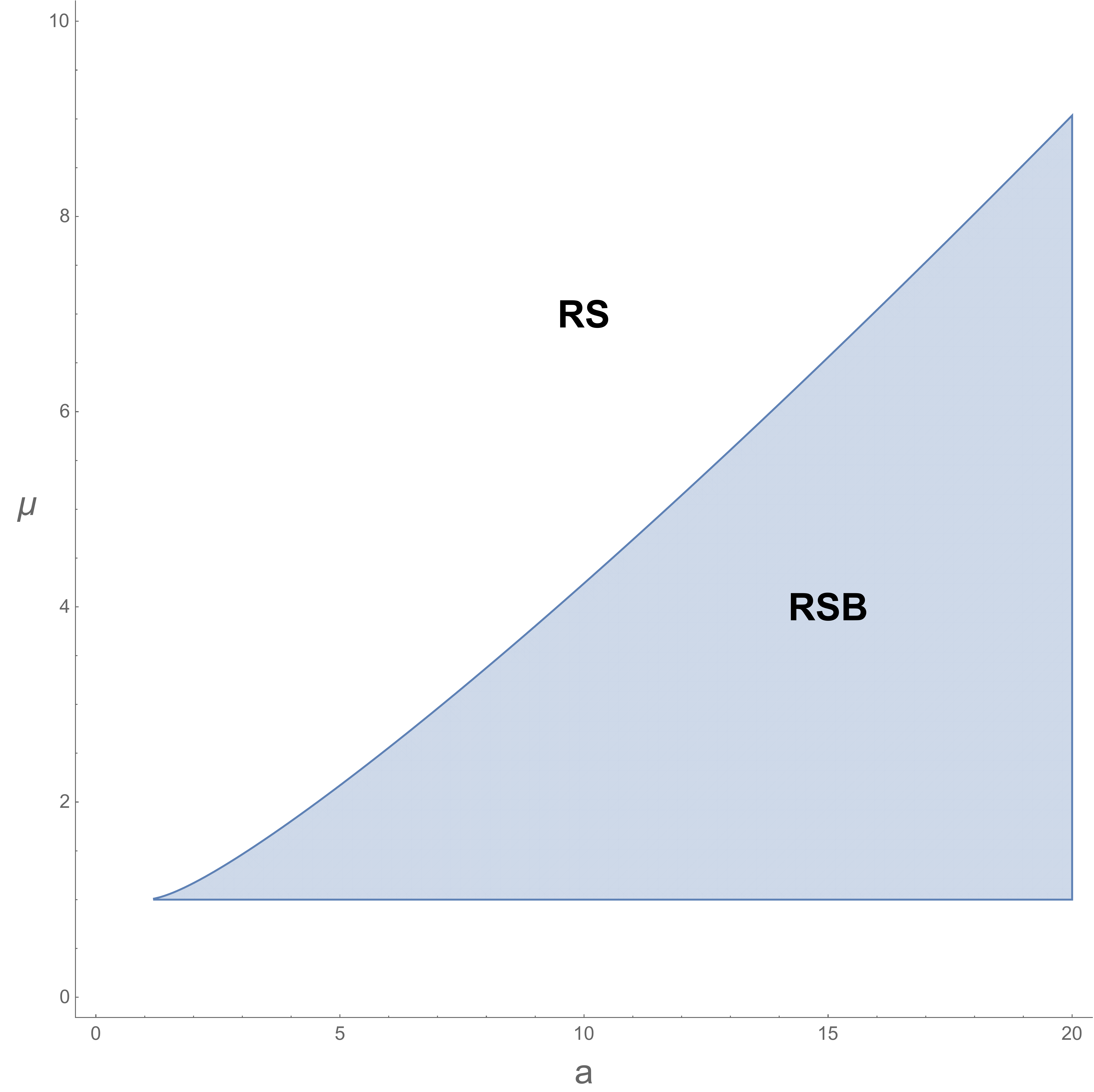}
  \caption{\label{figure1} Schematic Phase diagram in $(a,\mu)$ plane. In the shaded region of parameters
   $1<\mu<\frac{(a^{2/3}-a^{1/3}+1)^3}{a}$ replica symmetry can be broken for some amplitude of the noise.}
\end{figure}

 In contrast, at a fixed nonlinearity $a>1$ and not too big redundancy values  $\mu\in(1,\mu_{AT}(a))$  there exists generically an interval of scaled NSR's $\tilde{\gamma}^{(1)}_{AT}<\tilde{\gamma}<\tilde{\gamma}^{(2)}_{AT}$ such that the replica-symmetry is broken inside and preserved for $\tilde{\gamma}$ outside that interval. The exact values $\tilde{\gamma}_{AT}^{(1,2)}$ can be in general found only by numerically solving the  4th-order polynomial equation, see Eq.(\ref{p-RS-linquad}). At the same time, using that for large enough scaled NSR $\tilde{\gamma}>\tilde{\gamma}^{(2)}_{AT}$ the replica symmetry is restored, one can employ the RS equation  Eq.(\ref{p-RS-linquad}) to determine the behaviour of the quality parameter $p_{\infty}(\tilde{\gamma})$ as $\tilde{\gamma}\to \infty$. One finds that in all cases but one this quantity vanishes for asymptotically large values of NSR as  $p_{\infty}\sim \tilde{\gamma}^{-1/2}$, see Eq.(\ref{largeSNR-RS}), i.e. in qualitatively the same way as for purely linear system with $a=0$.

The only exceptional case, showing qualitatively different behaviour to the above picture, is that of {\it purely quadratic} encryption with vanishing linear component\footnote{It is worth noting that in the absence of linear component the encryption mapping $V({\bf s})$ in  Eq.(\ref{linearquadratic}) becomes invariant with respect to the reflections ${\bf s}\to -{\bf s}$. As a result, the least-square reconstruction may formally return solutions with negative values of the parameter $p_N$ in Eq.(\ref{qualirec}). To avoid this we consider the pure quadratic case as the limit $J_1\to 0$ taken {\it after} $N\to \infty$, which is enough to break the mentioned invariance.}, when $J_1\to 0$ at a fixed value of $J_2>0$. The appropriately rescaled NSR in this case is $\hat{\gamma}=\frac{\gamma}{J_2^2R^2}$.
In this limit the second threshold $\tilde{\gamma}^{(2)}_{AT}$ escapes to infinity and the replica symmetry is broken for {\it all} $\hat{\gamma}>\hat{\gamma}_{AT}=(\mu-1)^2/\mu$. Moreover, most importantly there exists a threshold NSR value $\hat{\gamma}_c=\mu-\frac{1}{2}>\hat{\gamma}_{AT}$ such that $p_{\infty}=0$ for $\hat{\gamma}>\hat{\gamma}_c$ making the reconstruction impossible. The full curve $p_{\infty}(\hat{\gamma})$ can be explicitly described in this case analytically.
In particular,  the behaviour close to the threshold NSR is given by $p_{\infty}\sim (\hat{\gamma}_c-\hat{\gamma})^{3/4}$ and
the non-trivial exponent $3/4$ is fully controlled by the replica symmetry breaking mechanism.

The existence of a sharp NSR threshold $\hat{\gamma}_c$ in the pure quadratic encryption case may have useful consequences
for security of transmitting the encrypted signal. Indeed, it is a quite common assumption that an eavesdropper may get access to
the transmitted signal by a channel with inferior quality, characterized by higher level of noise. This may
then result in impossibility for eavesdroppers to reconstruct the quadratically encoded signal even if the encoding algorithm is perfectly known to them.

\subsubsection{General remarks on the method}
The task of optimizing various random 'cost functions', not unlike ${\cal H}_{\bf s}({\bf x})$ in Eq.(\ref{Hamilt}),
is long known to be facilitated by a recourse to the methods of statistical mechanics, see e.g. \cite{Parisilec04} and \cite{MPV} for early references and introduction to the method. In that framework one encounters  the task of evaluating expectated values over distributions of random variables coming through the cost function in both numerator and denominator of the equations describing the quantities of interest, see the right-hand side of Eq.(\ref{disttermav}) below. Performing such averaging is known to be one of the central technical problems in the theory of disordered systems. One of the most powerful, though non-rigorous, methods of dealing with this problem at the level of theoretical physics is the (in)famous replica trick, see \cite{MPV} and references therein.  A considerable progress achieved in the last decades in developing rigorous aspects of that theory \cite{Bovier,Panchbook} makes this task, in principle, feasible for the cases when the random energy function ${\cal H}_{\bf s}({\bf x})$ is {\it Gaussian}-distributed.
  The model where configurations are restricted to the surface of a sphere are known in the spin-glass literature as
  'spherical models', but their successful treatment, originally nonrigorous \cite{CD1,CS,KTJ} and in recent years rigorous\cite{Auf2,Baik2016,Chen2017a,Chen2017b,Tal2006,Subag2017}, seems again be restricted to the normally-distributed case. In the present case however the cost function is {\it per se} not Gaussian, but represented as a sum of {\it squared} Gaussian-distributed terms. We are not aware of any systematic treatment of spherical spin glass models with such type of spin interaction. Some results obtained by extending replica trick treatment to this type of random functions were given by the present author in \cite{my}, but details were never published. To present the corresponding method on a meaningful example is one of the goals of the present paper. Indeed, we shall see that, with due modifications, the method is very efficient, and, when combined with the Parisi replica symmetry breaking Ansatz allows to get a reasonably detailed insight into the reconstruction problem.
  As squared gaussian-ditributed terms are common to many optimization problems based on the Least Square method, one may hope that the approach
  proposed in the present paper may prove to be of wider utility. In particular, an interesting direction of future research may be
  study of the minima, saddles and other structures of this type in the arising 'optimization landscape' following
  an impressive recent progress in this direction for Gaussian spherical model, see  \cite{RBBC2018}
 and references therein. This may help to devise better search algorithms for solutions of the optimization problems of this type.

  Another technical aspect of our treatment which is worth mentioning is as follows. In
  problems of this sort replica treatment is much facilitated by noticing  that after performing the disorder averaging the replicated partition function possesses a high degree of invariance: an arbitrary simultaneous
$O(N)$ rotation of all $n$ replica vectors ${\bf x}_a, \, a=1,\ldots,n$ leaves the integrand invariant.
 To exploit such an invariance in the most efficient way one may use a method suggested
 in the framework of Random Matrix Theory in the works \cite{F2002a,F2002b} \footnote{ Equivalent
 transformations were also suggested earlier in \cite{Percus87}, see also \cite{DDG1993}.}
 That method allowed one to convert the integrals over $N-$ component vectors ${\bf x}_{a},\, a=1,\ldots,n$ to a single positive-definite $n\times n$ matrix $Q_{ab}\ge 0$. Such transformation than
allows to represent the integrand in a form ideally suited for extracting the large-$N$ asymptotic of the integral.
In the context of spin glasses and related statistical mechanics systems this method was first successfully used in \cite{FS2007} and then \cite{FB2008b,FLD2013}, and most recently in \cite{KMZ2016}, and proved to be
a very efficient framework for implementing the Parisi scheme of replica symmetry breaking.
In the present problem however  the integrand has lesser invariance due to presence of a fixed direction exemplified by the original message ${\bf s}$. Namely, it is invariant only  with respect to rotations forming a subgroup of $O(N)$ consisting of all $N\times N$ orthogonal transformations $O_{\bf s}$ satisfying $O_{\bf s}^{T}O_{\bf s}={\bf 1}_N$ and $O_{\bf s}{\bf s}={\bf s}$.
 In the Appendix C we prove a Theorem which is instrumental in adjusting our approach to
 the present case of a fixed direction. One may hope that this generalization may have other applications beyond the present problem.

{\bf Acknowledgements}. The author is grateful to Jean-Philippe Bouchaud, Christian Schmidt, Guilhem Semerjian, Nicolas Sourlas and Francesco Zamponi for enlightening discussions and encouraging interest in this work, and to Dr. Mihail Poplavskyi for his help with
analysis of Eq.(\ref{AT-line-linquad}) and preparing figures for this article.\\
The financial support by  EPSRC grant  EP/N009436/1 "The many faces of random characteristic polynomials" is acknowledged with thanks.

\section{Statistical Mechanics approach to reconstruction problem}

\subsection{\sf General setting of the problem}

To put the least square minimization problem (\ref{leastsquare}) in the context of Statistical Mechanics, one follows the standard route   and interprets the cost function in Eq.(\ref{Hamilt}) as an energy associated with a configuration ${\bf x}^T$ of $N$ spin variables $(x_1,\ldots, x_N)$, constrained to the sphere of radius $|{\bf x}|=N\sqrt{R}$.
This allows one to treat our minimization problem as a problem of Statistical Mechanics, by introducing the temperature parameter $T>0$, and considering the Boltzmann-Gibbs weights $\pi_{\beta}({\bf x})={\cal Z}_{\beta}^{-1} e^{-\beta {\cal H}_{\bf s}({\bf x})}$ associated with any configuration
${\bf x}$ on the sphere, with ${\cal Z}_{\beta}$ being the partition function of the model for the inverse temperature $\beta=T^{-1}$ :
\be\label{6}
 {\cal Z}_{\beta}=\int_{|{\bf x}|=R\sqrt{N}} e^{-\beta {\cal H}_{\bf s}({\bf x})} d{\bf x}, \quad  d{\bf x}=\prod_{i=1}^N dx_i, \quad \,,
\ee
The power of the method is that in the zero-temperature limit
 $\beta\to \infty$ the Boltzmann-Gibbs weights concentrate on the set of globally minimal values of the cost function, so that
 for any well-behaving function $g({\bf x})$ the thermal average value $g_{\beta}({\bf x}):=\int g({\bf x}) \pi_{\beta}({\bf x})d{\bf x}$
 should tend to the value of that function evaluated at the argument corresponding to solutions of the minimization problem
 (\ref{leastsquare}). To this end we introduce the thermal average  $p^{(\beta)}_N$ of the distance function defined
in eq.(\ref{dist}) and consider its expected value  with respect to both the set of random functions $V_k({\bf x})$
and the noise ${\bf b}$:
\begin{equation}\label{disttermav}
\left\langle p^{(\beta)}_N\right\rangle:=\left\langle \frac{1}{{\cal Z}_{\beta}}\int_{|{\bf x}|=R\sqrt{N}}{\small \frac{({\bf x},{\bf s})}{NR^2}}\, e^{-\beta {\cal H}_{\bf s}({\bf x})} d{\bf x}\right\rangle_{V,{\bf b}}
\end{equation}
Our goal is to evaluate the above quantity for finite $\beta=1/T$ in the limit of large $N\gg 1$, and eventually perform the zero temperature limit $T\to 0$ thus extracting $ p_{\infty}:=\lim_{\beta\to \infty}\lim_{N\to \infty}\left\langle p^{(\beta)}_N\right\rangle$ providing us with a measure of the quality of the asymptotic signal reconstruction in the original optimization problem.

\subsection{\sf Replica trick}

In the framework of the replica trick one represents the
normalization factor ${\cal Z}^{-1}_{\beta}$ in the Boltzmann-Gibbs weights
formally as $1/{\cal Z}_{\beta}=\lim_{n\to 0}{\cal Z}_{\beta}^{n-1}$
and treats the parameter $n$ before the limit as a positive integer. This allows to rewrite
(\ref{disttermav}) formally as
\be\label{replim}
\left\langle p^{(\beta)}_N\right\rangle =\lim_{n\to 0}\left\langle p^{(\beta)}_{N,n}\right\rangle
\ee
where we defined
\begin{equation}\label{disttermavrep1}
\fl \left\langle p^{(\beta)}_{N,n}\right\rangle=\int_{|{\bf x_1}|=R\sqrt{N}}\ldots \int_{|{\bf x}_n|=R\sqrt{N}}\left[\frac{1}{n}\sum_{c=1}^n\frac{({\bf x}_c,\,{\bf s})}{NR^2}\right] \left\langle e^{-\beta \sum_{a=1}^n{\cal H}_{\bf s}({\bf x}_a)} \right\rangle\, \prod_{a=1}^n  d{\bf x}_a
\end{equation}
The disorder average can be now performed in the following steps. First, using the additive form of the cost function in Eq.(\ref{Hamilt})
and independence of ${\cal H}^{(k)}_{\bf s}({\bf x}_a)$ for different $k=1,\ldots,M$ we obviously have
\be \label{Mdecouple}
\left\langle e^{-\beta \sum_{a=1}^n{\cal H}_{\bf s}({\bf x}_a)} \right\rangle=\left\langle e^{-\beta \sum_{a=1}^n{\cal H}^{(k)}_{\bf s}({\bf x}_a)} \right\rangle^M
\ee
Using the Gaussian nature of $V({\bf x})$ entering to ${\cal H}^{(k)}_{\bf s}({\bf x})$ in a squared form, see Eq.(\ref{Hamilt}),
and exploiting the covariance structure (\ref{cov})  one can show that
\be\label{multichi}
\left\langle e^{-\beta \sum_{a=1}^n{\cal H}^{(k)}_{\bf s}({\bf x}_a)} \right\rangle
=\left[\det{{\cal G}({\bf x}_1,\ldots,{\bf x}_n;\,{\bf s})}\right]^{-1/2}
\ee
where we introduce the (positive definite) $n\times n$ matrix ${\cal G}({\bf x}_1,\ldots,{\bf x}_n;\,{\bf s})$ with entries
 \be\label{G}
\fl {\cal G}_{ab}({\bf x}_1,\ldots,{\bf x}_n;\,{\bf s})=\delta_{ab}+\beta\left[\sigma^2+\Phi(R^2)+\Phi\left(\frac{({\bf x}_a,{\bf x}_b)}{N}\right)-\Phi\left(\frac{({\bf x}_a,{\bf s})}{N}\right)-\Phi\left(\frac{({\bf x}_b,{\bf s})}{N}\right)\right]
\ee
For convenience of the reader we provide a derivation of the formula Eq.(\ref{multichi}) in the Appendix B. Note that this result is well-known in the probability literature, see e.g.\cite{PermProc}. We see that
\begin{equation}\label{disttermavrep2}
\fl  \left\langle  p^{(\beta)}_{N,n}\right\rangle=\int_{|{\bf x_a}|=R\sqrt{N}}\left[\frac{1}{n}\sum_{c=1}^n\frac{({\bf x}_c,\,{\bf s})}{NR^2}\right] \left[\det{{\cal G}({\bf x}_1,\ldots,{\bf x}_n;\,{\bf s})}\right]^{-M/2} \prod_{a=1}^n  d{\bf x}_a
\end{equation}
At this step it is very helpful to notice that the integrand in  (\ref{disttermavrep2}) possesses a high degree of invariance.
Namely, consider all possible rotations around
the axis whose direction is given by the vector ${\bf s}$. Such rotations form a subgroup of $O(N)$ consisting of all $N\times N$ orthogonal transformations $O_{\bf s}$ satisfying $O_{\bf s}^{T}O_{\bf s}={\bf 1}_N$ and $O_{\bf s}{\bf s}={\bf s}$.
Then the integrand in (\ref{disttermavrep2}) remains invariant under a simultaneous change
${\bf x}_a\to O_{\bf s} {\bf x}_a$ for all $a=1,\ldots,n$.
 In the Appendix C we prove a Theorem which is instrumental for implementing our previous approach to similar problems \cite{FS2007} to the present case of somewhat lesser invariance. Not surprisingly, in such a case the integration needs to go not only over
 $n\times n$ matrix of scalar products  $Q_{ab}=({\bf x}_a,{\bf x}_b)\ge 0$, but also over an $n$-component vector ${\bf t}=(t_1,\ldots,t_n)\in \mathbb{R}^n$ of projections $t_a=({\bf x}_a,{\bf s})$. Applying the Theorem  and rescaling for convenience the integration variables $Q_{ab}\to NQ_{ab}$ and ${\bf t}\to \sqrt{N} {\bf t}$ we
 bring Eq.(\ref{disttermavrep3}) to the form
 \begin{equation}\label{disttermavrep3}
\fl \left\langle p^{(\beta)}_{N,n}\right\rangle=N^{\frac{n(N-n-1)}{2}}\,C^{(o)}_{N-1,n}
 \int_{D_n} \left[ \frac{1}{n}\sum_{c=1}^n \frac{t_c}{R} \right]  \left[\det{Q}\right]^{-\frac{(n+2)}{2}} e^{-\frac{N}{2}F(Q,{\bf t})} dQ d{\bf t}
\end{equation}
 where $C^{(o)}_{N,n}$ is defined in Eq.(\ref{const}),  the integration goes over the domain
 \begin{equation}\label{domain}
 D_n: \left(Q>0, \quad {\bf t}\in \mathbb{R}^n, \quad Q_{aa}+t_a^2=R^2, \, \forall a=1,\ldots, n \right)
 \end{equation}
 and we defined
 \begin{equation}\label{Frep}
 F(Q,{\bf t})=\mu \mbox{\small Tr}\ln{\left[{\bf 1}_n+\beta\,g(Q,{\bf t})\right]} -\mbox{\small Tr}\ln{Q}
 \end{equation}
 with $n\times n$ matrix $g(Q,{\bf t})$ characterized by its entries (cf. Eq.(\ref{G}))
\begin{equation}\label{g}
\fl  g_{ab}(Q,{\bf t})=\sigma^2+\Phi(R^2)+\Phi\left(Q_{ab}+t_at_b\right)-\Phi\left(Rt_a\right)-\Phi\left(Rt_b\right)
\end{equation}
So far our treatment of  $\left\langle p_{N,n}^{(\beta)}\right\rangle$ was exact for any positive integer values
$N$ and $n$ satisfying $N>n+1$ and involved no approximations.
Our goal is however to extract the leading behaviour of that object as $N\gg 1$ and allowing formally $n$ to take non-integer values to be able to perform the replica limit $n\to 0$.

\subsection{ \sf Variational problem in the framework of Parisi Ansatz}
Clearly, the form of the integrand in Eq.(\ref{disttermavrep3}) being proportional  to the factor $e^{-\frac{N}{2}F(Q,{\bf t})}$ is suggestive of using the Laplace (a.k.a saddle-point or steepest descent) method.
In following this route we resort to a non-rigorous and heuristic, but computationally efficient scheme of Parisi replica symmetry breaking \cite{MPV}.  We implement this scheme in a particular variant most natural for models with rotational invariance, going back to Crisanti and Sommers paper\cite{CS}, and somewhat better explained in the Appendix A of \cite{FS2007}, and in even more detail in the Appendix C of \cite{logmultlec}. We therefore won't discuss the method itself in the present paper, only giving a brief account of necessary steps.

The scheme starts with a standard assumption that in the replica limit $n\to 0$ the integral is dominated by configurations
of matrices $Q$ which for finite integer $n$ have a special hierarchically built
structure characterized by the sequence of integers
\begin{equation}\label{parisiseq}
n=m_0\ge m_1\ge m_2\ge \ldots\ge m_k\ge m_{k+1}=1
\end{equation}
and the values placed in the off-diagonal entries of the $Q$ matrix block-wise, and
satisfying:
\begin{equation}\label{parisiseq1}
0<q_0\le q_1\le q_2\le \ldots\le q_k
\end{equation}
Finally, we complete the procedure by filling in the $n$ diagonal
entries $Q_{aa}$ of the matrix $Q$ with one and the same
value $Q_{aa}=q_d:=q_{k+1}\ge q_k$. Note that in our particular case the diagonal entries $q_d$ must in fact be chosen in the form
$q_d=R^2-t_a^2$, in order to respect the constraints impose by the integration domain Eq.(\ref{domain}).
As to the vector ${\bf t}$ of  variables $t_{a}$, we are making an additional assumption that with respect to those
variables the integral is in fact dominated by equal values: $t_a=t, \forall a=1,\ldots,n$.

Obviously, the matrix $g(Q,{\bf t})$ defined in (\ref{g}) inherits the hierarchical structure from $Q$, with
parameters $m_l, \, l=0,k+1$ shared by both matrices, but parameters $q_l$ replaced by parameters $g_l$
given by
\begin{equation}\label{entries_g}
 g_l=\sigma^2+\Phi(R^2)-2\Phi(Rt)+\Phi(q_l+t^2)), \quad l=0,1,\ldots, k
\end{equation}
and
\[
 g_{k+1}=\sigma^2+2\Phi(R^2)-2\Phi(Rt):=g_d
\]

The next task of the scheme is to express both ${\small Tr}\ln{g(Q,{\bf t})}$ and $\mbox{\small Tr}\ln{Q}$ in terms of
the parameters entering Eqs.(\ref{parisiseq}) and (\ref{parisiseq1}).  This is most easily achieved by writing down
all distinct eigenvalues $\lambda_1,\ldots,\lambda_{k+2}$ of the involved matrices, and their degeneracies
$d_{i+2}=n\left(m^{-1}_{i+1}-m^{-1}_{i}\right)\, \forall i=0,\ldots,k$, and $d_1=1$.
 For the matrix $Q$ those eigenvalues are listed, e.g.,
in the appendix C of \cite{logmultlec}, and for the matrix $g(Q,{\bf t})$ the corresponding expressions can be obtained
from those for $Q$ by replacing $q_l$ by parameters $g_l$ from Eq.(\ref{entries_g}).
The subsequent treatment is much facilitated by
introducing the following (generalized) function of the variable
$q$:
\begin{equation}\label{xstep}
x(q)=n+\sum_{l=0}^k (m_{l+1}-m_l)\,\theta(q-q_l)
\end{equation}
where we use the notation $\theta(z)$ for the Heaviside step
function: $\theta(z)=1$ for  $z>0$ and zero otherwise. In view of
the inequalities Eq.(\ref{parisiseq},\ref{parisiseq1}) the
function $x(q)$ is piecewise-constant non-increasing, and changes
between $x(0<q<q_0)=m_0\equiv n$ through $x(q_{i-1}<q<q_i)=m_i$ for $i=1, \ldots,k$ to finally $x(q_k<q<q_d)=m_{k+1}\equiv 1$. A clever observation by
Crisanti and Sommers allows one to express eigenvalues of any function of the hierarchical matrix Q in terms of simple integrals
involving $x(q)$. In particular, for eigenvalues $\lambda^{(g)}_l$ of the matrix $g(Q,{\bf t})$ we have:
\begin{equation}\label{eigen_g}
\lambda^{(g)}_{1}=\int_{0}^{q_d}x(q)\frac{dg_t}{dq}\,dq,\quad \mbox{and}\quad \lambda^{(g)}_{i+2}=\int_{q_i}^{q_d}x(q)\frac{dg_t}{dq}\,dq, \quad i=0, \ldots, k
\end{equation}
where we introduced a piecewise-continuous function $g_t(q), q\in[0,q_d]$ such that in the interval $q\in[q_0,q_k]$ it is given by
\begin{equation}\label{func_gA}
g_t(q_{0}\le q\le q_{k})=\sigma^2+\Phi(R^2)-2\Phi(Rt)+\Phi(q+t^2))
\end{equation}
whereas outside that interval it has two constant values:
\begin{equation}\label{func_gB}
 \fl g_t(0\le q<q_0)=0 \quad \mbox{ and} \quad  g_t(q_{k}< q\le q_d:=q_{k+1})=\sigma^2+2\Phi(R^2)-2\Phi(Rt)
\end{equation}
In particular, $\lambda^{(g)}_{1}$ can be further rewritten as
\begin{equation}\label{eigen_g_1}
\fl \lambda^{(g)}_{1}=\int_{0}^{q_0}x(q)\frac{dg_t}{dq}\,dq+\int_{q_0}^{q_d}x(q)\frac{dg_t}{dq}\,dq=ng_t(q_0)+\int_{q_0}^{q_d}x(q)\frac{dg_t}{dq}\,dq
\end{equation}
Such a representation, together with the definition Eq.(\ref{xstep}) of the function $x(q)$ facilitates calculating quantities interesting to us in the replica limit as:
\begin{equation}\label{replica_lng}
\fl \lim_{n\to 0}\frac{1}{n}\mbox{\small Tr}\ln{\left({\bf 1}_n+\beta\,g(Q,{\bf t})\right)}
=\lim_{n\to 0}\frac{1}{n}
\left[\ln{\left(1+\beta \lambda^{(g)}_1\right)}+ \sum_{i=0}^k d_{i+2} \ln{\left(1+\beta \lambda^{(g)}_{i+2}\right)}\right]
\end{equation}
\[
=lim_{n\to 0}\frac{1}{n}
\left[\ln{\left(1+\beta n g_t(q_0)+\beta\int_{q_0}^{q_d}x(q)\frac{dg_t}{dq}\,dq\right)}\right.
\]
\[
\left.+\int_{q_0-0}^{q_k+0}\ln{\left(1+\beta \int_{q}^{q_d}x(q)\frac{dg_t}{dq}\,dq \right)}\frac{d}{dq}\left[\frac{1}{x(q)}\right]\,dq\right]
\]
In the last term it is convenient to integrate by parts, and use $x(q_k+0)=1$ and $x(q_0-0)=n$, which after
obvious regrouping of terms reduces the right-hand side of Eq.(\ref{replica_lng}) as
\begin{equation}\label{replica_lngA}
\fl \lim_{n\to 0}\frac{1}{n}
\left[\ln{\left(1+\beta n g_t(q_0)+\beta\int_{q_0}^{q_d}x(q)\,g'_t(q)\,dq\right)}-\ln{\left(1 +\beta\int_{q_0}^{q_d}x(q)g_t'(q)\,dq\right)}\right]
\end{equation}
\[
+\lim_{n\to 0}\left\{\ln{\left[1+\beta \left(g_t(q_d)-g_t(q_k)\right)\right]}+\beta\int_{q_0}^{q_k}\frac{g_t'(q)dq}
{1 +\beta\int_{q}^{q_d}x(\tilde{q})g'_t(\tilde{q})d\tilde{q}}\right\}
\]
where we denoted $g_t'(q):=\frac{dg_t}{dq}$. The limit $n\to 0$ is now easy to perform following
the general prescription of the Parisi method: in such a limit the
inequality Eq.(\ref{parisiseq}) should be reversed:
\begin{equation}\label{parisiseq2a}
n=0\le m_1\le m_2\le \ldots\le m_k\le m_{k+1}=1
\end{equation}
and the function $x(q)$ is now transformed to a non-decreasing
function of the variable $q$ in the interval $q_0\le q \le q_k$,
and satisfying outside that interval the following properties
\begin{equation}\label{outsidea}
x(q<q_0)=0,\quad \mbox {and}\quad  x(q>q_k)=1.
\end{equation}
In general, such a function also depends on the increasing sequence
of $k$ real parameters $m_l$ described in Eq.(\ref{parisiseq2a}) .
Performing the corresponding limit and taking into account that in view of $x(q<q_0))=0$ we have
\[
\int_{0}^{q_0}\frac{g_t'(q)dq}
{1 +\beta\int_{q}^{q_d}x(\tilde{q})g'_t(\tilde{q})d\tilde{q}}=\frac{g_t(q_0)}
{1 +\beta\int_{q_0}^{q_d}x(q)g'_t(q)\,d q}
\]
we eventually see that
\begin{equation}\label{replica_lngB}
\fl \lim_{n\to 0}\frac{1}{n}\mbox{\small Tr}\ln{\left({\bf 1}_n+\beta\,g(Q,{\bf t})\right)}=\ln{\left[1+\beta \left(g_t(q_d)-g_t(q_k)\right)\right]}+\beta\int_{0}^{q_k}\frac{g_t'(q)dq}
{1 +\beta\int_{q}^{q_d}x(\tilde{q})g'_t(\tilde{q})d\tilde{q}}
\end{equation}
and by a similar calculation also find:
\begin{equation}\label{replica_lngC}
 \lim_{n\to 0}\frac{1}{n}\mbox{\small Tr}\ln{Q}=\ln{\left(q_d-q_k\right)}+\int_{0}^{q_k}\frac{dq}
{\int_{q}^{q_d}x(\tilde{q})d\tilde{q}}
\end{equation}
The two formulas Eq.(\ref{replica_lngB})-(\ref{replica_lngC}) provide us therefore with a full formal control of the main exponential factor $e^{-\frac{N}{2}F(Q,{\bf t})}$ in Eq.(\ref{disttermavrep3}) for $N\to \infty$ in the replica limit $n\to 0$.
Note that the fact that the limit in the left hand-side of Eq.(\ref{replica_lngC}) is finite implies also that
$\det{Q}|_{n\to 0}\to 1$ further implying
 $\left[\det{Q}\right]^{-\frac{(n+2)}{2}} |_{n\to 0}=1$ in Eq.(\ref{disttermavrep3}). Collecting finally all factors in Eq.(\ref{disttermavrep3}) when performing the replica limit $n\to 0$, using explicit forms Eq.(\ref{func_gA})-(\ref{func_gB}),
 remembering $q_d=R^2-t^2$ and finally understanding by $x(q)$ only its non-trivial part in the interval $q_0,q_k$
 we arrive to the following \\
 {\bf Proposition 2.1}:
 {\it Given the values of real parameters $R>0$ and
 $\mu>1$,  consider the functional
\begin{equation}\label{mainfiniteT}
\fl \phi[x(q);q_0,q_k,t]=\mu\ln{\left(1+\beta\left[\Phi(R^2)-\Phi(q_k+t^2)\right]\right)}-\ln{(R^2-t^2-q_k)}
\end{equation}
\[
\fl +\mu\frac{\beta\left[\sigma^2+\Phi(R^2)-2\Phi(Rt)+\Phi(q_0+t^2)\right]}
{1+\beta\left[\Phi(R^2)-\Phi(q_k+t^2)\right]+\beta\int_{q_0}^{q_k}x(u)\Phi'(u+t^2)du}-\frac{q_0}{R^2-t^2-q_k+\int_{q_0}^{q_k}x(u)du}
\]
\[
\fl +\mu\beta\int_{q_0}^{q_k}\frac{\Phi'(q+t^2)\,dq}
{1+\beta\left[\Phi(R^2)-\Phi(q_k+t^2)\right]+\beta\int_{q}^{q_k}x(u)\Phi'(u+t^2)du}-
\int_{q_0}^{q_k}\frac{dq}{R^2-t^2-q_k+\int_{q}^{q_k}x(u)du}
\]
  which depends on the parameters $-R\le t\le R$,  and $0\le q_0\le q_k<q_d=R^2-t^2$ and a non-decreasing
function $x(q)$ of the variable $q$ in the interval $q_0\le q \le q_k$.
Then in the framework of the replica trick the asymptotic  mean value of the quality parameter $\left\langle p^{(\beta)}_N\right\rangle$ as $N\to \infty$  is given by
\be\label{replimParisi}
\lim_{N\to \infty}\left\langle p^{(\beta)}_N\right\rangle = \frac{t}{R}
\ee
 where the specific value of the parameter $t$ is found by simultaneously minimizing the functional $\phi[x(q);q_0,q_k,t]$ over $t$ and
 maximizing it over all other parameters and the function $x(q)$.
 }

 Recall, however, that for the purposes of our main goal the quantity  $\left\langle p^{(\beta)}_N\right\rangle$
  is only of auxiliary interest, and is used to provide an access to its 'zero-temperature' limit $\beta=\frac{1}{T}\to \infty$ which is expected to coincide with the quality parameter characterizing the performance of our signal reconstruction scheme. A simple inspection shows that in such a limit the combination  $T\phi[x(q);q_0,q_k,t]$ does have a well-defined finite value if we make the following {\it low temperature Ansatz}
  valid for $T\to 0$
  \begin{equation}\label{LTA}
  q_k=R^2-t^2-v\,T, \quad q_0=R^2-t^2-Q, \quad \beta x(u)\to w(u):=w_s(u+t^2)
  \end{equation}
   with $v,Q$ and $w_s(u)$ tending to a well-defined finite limit as $T\to 0$. Performing the corresponding limit in Eq.(\ref{mainfiniteT}) and changing $u\to u-t^2$ one arrives at the statement of the {\bf Proposition 1} in the Main Results section.

\section{Analysis of the variational problem}
To solve the arising problem of extremizing the functional ${\cal E}[w_s(u);Q,v,t]$ from Eq.(\ref{mainzerotemp}) we first consider the stationarity equations with respect to three parameters: $t,Q$ and $v$. The conditions $\frac{\partial\mathcal{E}}{\partial t}=0$
and $\frac{\partial\mathcal{E}}{\partial Q}=0$ yield the two equations, the first being
\begin{equation}\label{stat-t}
\frac{t}{v+\int_{R^2-Q}^{R^2}w_s(u)\,du}=\mu\,\frac{R\Phi'(Rt)}{1+v\Phi'(R^2)+\int_{R^2-Q}^{R^2}\,w_s(u)\Phi'(u)\,du}
\end{equation}
and, assuming that $w_{s}(R^2-Q)\ne 0$, the second one:
\begin{equation}\label{stat-Q}
\fl \frac{R^2-t^2-Q}{\left(v+\int_{R^2-Q}^{R^2}w_s(u)\,du\right)^2}=\mu\,\Phi'(R^2-Q)\,\frac{\sigma^2+\Phi(R^2)-2\Phi(Rt)+\Phi(R^2-Q)}
{\left(1+v\Phi'(R^2)+\int_{R^2-Q}^{R^2}\,w_s(u)\Phi'(u)\,du\right)^2}
\end{equation}
The Eq.(\ref{stat-Q}) can be used to simplify the third equation arising from the stationarity  condition $\frac{\partial\mathcal{E}}{\partial v}=0$ bringing it eventually to the following form:
\begin{equation}\label{stat-v}
\fl \frac{R^2-t^2-Q}{\left(v+\int_{R^2-Q}^{R^2}w_s(u)\,du\right)^2}\left[1-\frac{\Phi'(R^2)}{\Phi'(R^2-Q)}\right]
+\int_{R^2-Q}^{R^2}\frac{dq}{\left(v+\int_{q}^{R^2}w_s(u)\,du\right)^2}
\end{equation}
\[
=\mu\,\Phi'(R^2)\,\int_{R^2-Q}^{R^2}\frac{\Phi'(q)dq}
{\left(1+v\Phi'(R^2)+\int_{q}^{R^2}\,w_s(u)\Phi'(u)\,du\right)^2}
\]

\subsection{\sf Replica symmetric solution}

Notice that the last equation Eq.(\ref{stat-v})
is identically satisfied with the choice $Q=0$, defining the so-called Replica Symmetric
(RS) solution. For such a choice the interval $[R^2-Q,R^2]$ of the support of the function $w_s(u)$ shrinks to zero, making
that function immaterial for the variational procedure. Moreover, the equations (\ref{stat-t})-(\ref{stat-Q}) drastically
simplify yielding the pair:
\begin{equation}\label{stat-RS}
 \frac{t}{v}=\mu\,\frac{R\Phi'(Rt)}{1+v\Phi'(R^2)}, \quad \frac{R^2-t^2}{v^2}=\mu\,\Phi'(R^2)\,\frac{\sigma^2+2\Phi(R^2)-2\Phi(Rt)}
{\left(1+v\Phi'(R^2)\right)^2}
\end{equation}
Remarkably, this pair can be further reduced to a single equation in the variable $p=t/R$, precisely
one given in the list of Main Results, see Eq.(\ref{p-RS-Main}), thus providing the asymptotic value of the quality reconstruction parameter $p_{\infty}=p$.

As expected, in the case of no noise $\gamma=0$ the solution of Eq.(\ref{p-RS-Main}) is provided by $p=1$ corresponding to the perfect reconstruction of the source signal. It is also easy to treat the equation perturbatively in the case of weak noise $\gamma\to 0$
and obtain to the leading order:
\begin{equation}\label{p-RS_pert}
p=1-\frac{\gamma}{2(\mu-1)\Phi'(R^2)}+o(\gamma)
\end{equation}
In particular, this result agrees with the first-order perturbation theory analysis for the linear case $\Phi(u)=u$, see Eq.(\ref{qualiasylin}) and Appendix A,
and generalizes it to a generic nonlinearity. It also emphasizes the natural fact that the signal recovery becomes very sensitive to the noise for the values of Encryption Redundancy Parameter $\mu\to 1$.

For the linear-quadratic family Eq.(\ref{linearquadratic}) with the covariance structure of the form $\Phi\left(u\right)=J_1^2\, u+\frac{1}{2}J_{2}^2u^2$ the equation Eq.(\ref{p-RS-Main}) can be readily studied non-perturbatively for any value of NSR.
We start with two limiting cases in the family: that of 'purely linear' and 'purely quadratic' encryptions.
In the former case $J_2=0$, but $J_1\ne 0$ and after introducing the scaled NSR $\tilde{\gamma}:=\gamma/J^2_1$ we arrive at a cubic equation
 \begin{equation}\label{p-RS-lin}
 p^2\left(\tilde{\gamma}+2(1-p)\right)=\mu (1-p^2)
\end{equation}
In particular, $p$ is nonvanishing for any value of the scaled NSR $0\le \tilde{\gamma}<\infty$, and tends to zero as $p\sim \sqrt{\mu/\tilde{\gamma}}$
for $\tilde{\gamma}\gg 1$, in full agreement with the direct perturbation theory approach, see appendix A. For intermediate NSR values the solution can be easily plotted numerically, see Fig.2.

 \begin{figure}[h!]
 \centering
  \includegraphics[width=0.6\textwidth]{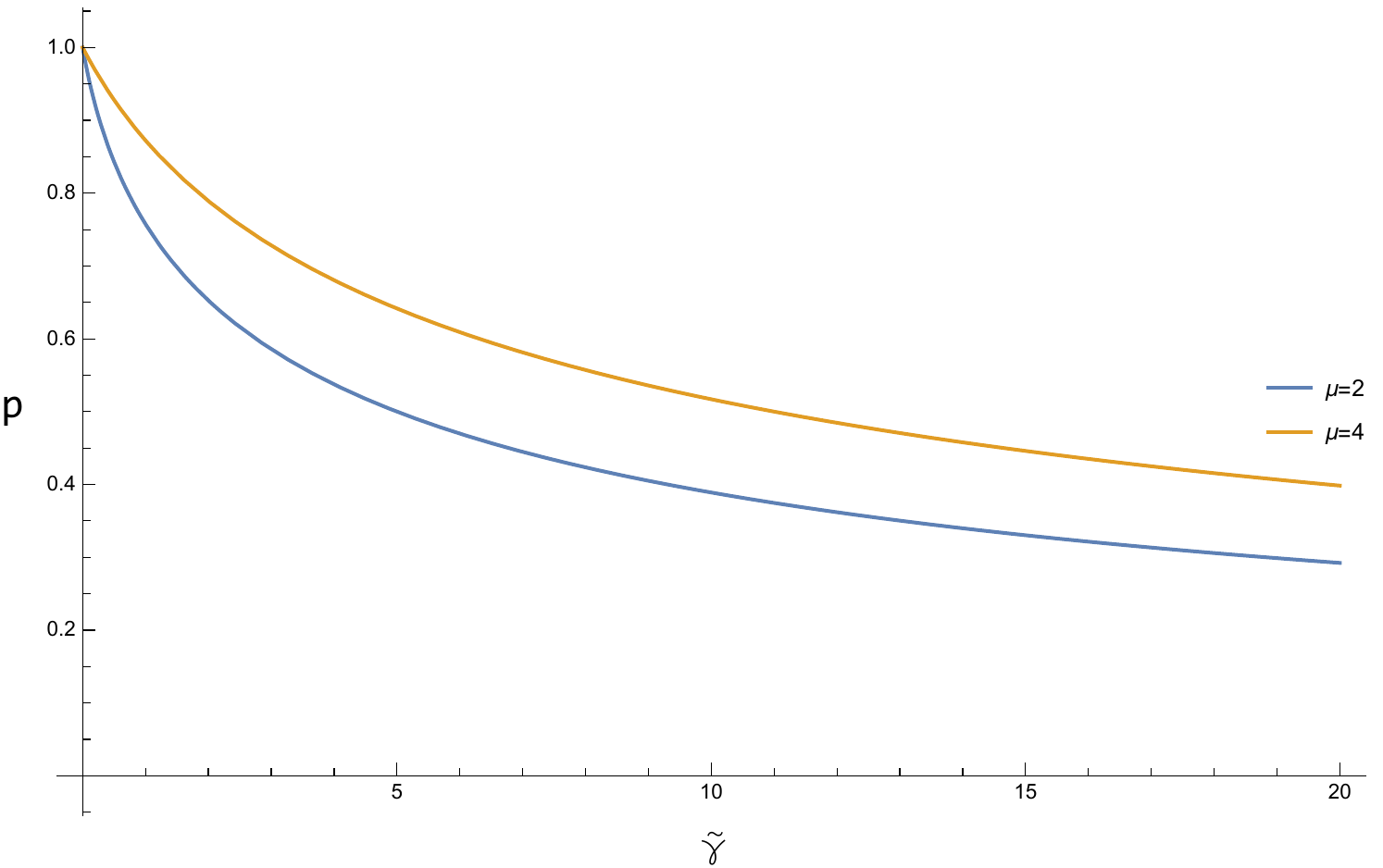}
  \caption{\label{figure1} The quality parameter $p$ as a function of the scaled noise-to-signal ratio $\tilde{\gamma}$ for purely linear encryptions and two different values of the Encryption Redundancy Parameter, $\mu=2$ and $\mu=4$.}
\end{figure}

In the opposite case of purely quadratic encryption when $J_1=0$ but $J_2\ne 0$ the equation Eq.(\ref{p-RS-Main}) is biquadratic, so that one can find the RS solution explicitly. Introducing the rescaled NSR pertinent to this limit as $\hat{\gamma}:=\frac{\gamma}{J_2^2R^2}$ we have
\begin{equation}\label{p-RS-quad}
 p=\left\{\begin{array}{cc} \sqrt{1-\frac{\hat{\gamma}}{\mu-1}} &  \mbox{ if}\,\, \hat{\gamma}\le \hat{\gamma}^{(RS)}_c:=\mu-1\\
 0 & \hat{\gamma} >\hat{\gamma}^{(RS)}_c
 \end{array} \right.
\end{equation}
Thus, in this case the replica-symmetric solution predicts the existence of a NSR threshold $\hat{\gamma}^{(RS)}_c = \mu-1$ beyond which meaningful reconstruction of the encrypted signal is impossible. We will see later on that although this conclusion is qualitatively correct, the actual value for the threshold and the critical exponent controlling the behaviour close to the threshold is different and is obtained when the phenomenon of the replica symmetry breaking is taken into account.

Finally, in the case of a generic linear-quadratic encryption with both $J_1\ne 0$ and $J_2\ne 0$ the resulting equation Eq.(\ref{p-RS-Main}) is a general polynomial of the fourth degree. Introducing again the scaled NSR $\tilde{\gamma}:=\gamma/J^2_1$ and  a parameter characterizing effective non-linearity of the mapping $a:=(R\,J_2/J_1)^2$ we can rewrite the equation as:
 \begin{equation}\label{p-RS-linquad}
(1+a) p^2\left(\tilde{\gamma}+2(1-p)+a(1-p^2)\right)=\mu (1-p^2)(1+ap)^2
\end{equation}
In particular, we see  that $p$ tends to zero as NSR $\tilde{\gamma}\to \infty$ as in the purely linear case:
\begin{equation}\label{largeSNR-RS}
p = \sqrt{\frac{\mu}{1+a} \frac{1}{\tilde{\gamma}}}+o(\tilde{\gamma}^{-1/2})
\end{equation}
We will see in the next section that generically for linear-quadratic encryptions with  big enough, but finite nonlinear component
$1<a<\infty$   the replica-symmetric solution of the variational problem is not correct in some interval of the scaled NSR $\tilde{\gamma}_{AT}^{(1)}<\tilde{\gamma}<\tilde{\gamma}_{AT}^{(2)}$, and should be replaced with one involving $Q\ne 0$. Nevertheless, asymptotic decay of the quality parameter $p$ for $\gamma\to \infty$
is always given by Eq.(\ref{largeSNR-RS}), apart from the only limiting case of purely quadratic encryption, when $a\to \infty$.

\subsection{\sf Solution with fully broken replica symmetry}
The goal of the present section is to seek for a solution of the variational problem for the functional Eq.(\ref{mainzerotemp}) which breaks the replica symmetry, so that
 $Q>0$. Doing this necessarily implies taking the function $w_s(u)$ into account, and deriving the equation involving such a function. The corresponding equation is obtained by requiring stationarity of the functional ${\cal E}$ with respect to a variation of
$w_s(u)$, assuming that function to be continuous in the interval $q\in[R^2-Q,R^2]$\footnote{Attempts to use the so-called 1RSB
scheme corresponding to a stepwise-discontinuous function $w_s(u)$ did not yield any viable solution respecting
the inequalities for the parameters involved.} . For every value of $q$ in that interval it yields the equation
\begin{equation}\label{stat-w}
\fl \frac{R^2-t^2-Q}{\left(v+\int_{R^2-Q}^{R^2}w_s(u)\,du\right)^2}
+\int_{R^2-Q}^{q}\frac{d\tilde{q}}{\left(v+\int_{\tilde{q}}^{R^2}w_s(u)\,du\right)^2}
\end{equation}
\[
\fl =\mu\,\Phi'(q)\left[\frac{\sigma^2+\Phi(R^2)-2\Phi(Rt)+\Phi(R^2-Q)}
 {\left(1+v\Phi'(R^2)+\int_{R^2-Q}^{R^2}\,w_s(u)\Phi'(u)\,du\right)^2}+\int_{R^2-Q}^{q}\frac{\Phi'(\tilde{q})d\tilde{q}}
{\left(1+v\Phi'(R^2)+\int_{\tilde{q}}^{R^2}\,w_s(u)\Phi'(u)\,du\right)^2}\right]
\]
which using again Eq.(\ref{stat-Q}) can be simplified into
\begin{equation}\label{stat-w1}
\fl \frac{R^2-t^2-Q}{\left(v+\int_{R^2-Q}^{R^2}w_s(u)\,du\right)^2}\left[1-\frac{\Phi'(q)}{\Phi'(R^2-Q)}\right]
+\int_{R^2-Q}^{q}\frac{d\tilde{q}}{\left(v+\int_{\tilde{q}}^{R^2}w_s(u)\,du\right)^2}
\end{equation}
\[
 =\mu\,\Phi'(q)\,\int_{R^2-Q}^{q}\frac{\Phi'(\tilde{q})d\tilde{q}}
{\left(1+v\Phi'(R^2)+\int_{\tilde{q}}^{R^2}\,w_s(u)\Phi'(u)\,du\right)^2}
\]
Our first observation is that setting $q=R^2$ in Eq.(\ref{stat-w1}) in fact reproduces  Eq.(\ref{stat-v}),
so the fundamental system comprises three rather than four independent stationarity conditions: Eq.(\ref{stat-t}), Eq.(\ref{stat-Q}) and either Eq.(\ref{stat-w}) or Eq.(\ref{stat-w1}) . Next we observe that Eq.(\ref{stat-t}) can be rewritten as
\begin{equation}\label{stat-t1}
\frac{1+v\Phi'(R^2)+\int_{R^2-Q}^{R^2}\,w_s(u)\Phi'(u)\,du}{v+\int_{R^2-Q}^{R^2}w_s(u)\,du}=\mu\, \frac{R\Phi'(Rt)}{t}
\end{equation}
which when substituted to Eq.(\ref{stat-Q}) yields the following equation
\begin{equation}\label{stat-Q1-fin}
\fl \mu\,R^2\left[\Phi'(Rt)\right]^2(R^2-t^2-Q)=t^2\Phi'(R^2-Q)\left[\sigma^2+\Phi(R^2)-2\Phi(Rt)+\Phi(R^2-Q)\right]
\end{equation}
After introducing the variable $p=t/R$, and the NSR $\gamma=\sigma^2/R^2$ the above equation is presented in the Main Results section, see (\ref{RSB-Main-1}).

At the next step we differentiate Eq.(\ref{stat-w}) over the variable $q$, and find that for any $q\in [R^2-Q,R^2]$ holds:
\begin{equation}\label{der-stat-w}
\frac{1}{\left(v+\int_{q}^{R^2}w_s(u)\,du\right)^2}=\mu\,
\frac{\left[\Phi'(q)\right]^2}{\left(1+v\Phi'(R^2)+\int_{q}^{R^2}\,w_s(u)\Phi'(u)\,du\right)^2}
\end{equation}
\[
\fl +\mu\,\Phi''(q)\left[\frac{\sigma^2+\Phi(R^2)-2\Phi(Rt)+\Phi(R^2-Q)}
 {\left(1+v\Phi'(R^2)+\int_{R^2-Q}^{R^2}\,w_s(u)\Phi'(u)\,du\right)^2}+\int_{R^2-Q}^{q}\frac{\Phi'(\tilde{q})d\tilde{q}}
{\left(1+v\Phi'(R^2)+\int_{\tilde{q}}^{R^2}\,w_s(u)\Phi'(u)\,du\right)^2}\right]
\]
Now, by comparing Eq.(\ref{der-stat-w}) with Eq.(\ref{stat-w}) and assuming that $\Phi''(q)\ne 0$
one arrives to the following relation:
\begin{equation}\label{der-stat-w1}
 \frac{R^2-t^2-Q}{\left(v+\int_{R^2-Q}^{R^2}w_s(u)\,du\right)^2}+\int_{R^2-Q}^{q}\frac{d\tilde{q}}
{\left(v+\int_{\tilde{q}}^{R^2}w_s(u)\,du\right)^2}
\end{equation}
\[
\fl =\frac{\Phi'(q)}{\Phi''(q)}\left[\frac{1}{\left(v+\int_{q}^{R^2}w_s(u)\,du\right)^2}-\mu\,\frac{\left[\Phi'(q)\right]^2}
{\left(1+v\Phi'(R^2)+\int_{q}^{R^2}\,w_s(u)\Phi'(u)\,du\right)^2}\right]
\]
We further substitute the value $q=R^2-Q$ in the above getting
\begin{equation}\label{der-stat-w2}
 \frac{R^2-t^2-Q}{\left(v+\int_{R^2-Q}^{R^2}w_s(u)\,du\right)^2}
\end{equation}
\[
\fl =\frac{\Phi'(R^2-Q)}{\Phi''(R^2-Q)}\left[\frac{1}{\left(v+\int_{R^2-Q}^{R^2}w_s(u)\,du\right)^2}-\mu\,\frac{\left[\Phi'(R^2-Q)\right]^2}
{\left(1+v\Phi'(R^2)+\int_{R^2-Q}^{R^2}\,w_s(u)\Phi'(u)\,du\right)^2}\right]
\]
which we further rearrange into
\begin{equation}\label{der-stat-w3}
\fl \frac{\left(1+v\Phi'(R^2)+\int_{R^2-Q}^{R^2}\,w_s(u)\Phi'(u)\,du\right)^2}{\left(v+\int_{R^2-Q}^{R^2}w_s(u)\,du\right)^2}
 \left[\frac{\Phi'(R^2-Q)}{\Phi''(R^2-Q)}-(R^2-t^2-Q)\right]
=\mu\, \frac{\left[\Phi'(R^2-Q)\right]^3}{\Phi''(R^2-Q)}\,.
\end{equation}
Finally, upon using Eq.(\ref{stat-t1}) the above relation is transformed into the following equation:
\begin{equation}\label{der-stat-fin}
\fl \left[\Phi'(R^2-Q)\right]^3\,\frac{t^2}{R^2}=\mu \left[\Phi'(Rt)\right]^2 \left[\Phi'(R^2-Q)-\Phi''(R^2-Q)(R^2-t^2-Q)\right]
\end{equation}
which is yet another equation presented in the Main Results section, see Eq.(\ref{RSB-Main-2}).

We therefore conclude that the pair of equations Eq.(\ref{stat-Q1-fin}) and Eq.(\ref{der-stat-fin})
is sufficient for finding the values of the parameters $t$ and $Q$, and hence for determining the value of $p$ giving the quality of the reconstruction procedure.

Using the above pair, the first task is to determine the range of NSR parameter $\gamma=\sigma^2/R^2$ where the solution with $Q>0$ is at all possible.
The boundary of this region which we denote as $\gamma_{AT}$ (in the general spin-glass context such boundaries are known as the de-Almeida-Thouless lines\cite{AT}) can be found by
setting $Q=0$ in  Eq.(\ref{stat-Q1-fin}) and Eq.(\ref{der-stat-fin}), yielding the system of two equations:
\begin{equation}\label{stat-Q1-fin-AT}
 \fl\mu\,R^2\left[\Phi'(Rt)\right]^2(R^2-t^2)=t^2\Phi'(R^2)\left[R^2\gamma_{AT}+\Phi(R^2)-2\Phi(Rt)+\Phi(R^2)\right]
\end{equation}
and
\begin{equation}\label{AT-line-gen}
\left[\Phi'(R^2)\right]^3\,p^2 = \mu \left[\Phi'(Rt)\right]^2 \left[\Phi'(R^2)-\Phi''(R^2)(R^2-t^2)\right]
\end{equation}
Moreover, it is not difficult to understand that by replacing in Eq.(\ref{AT-line-gen}) the equality sign $=$ with the inequality sign $\le$  defines the NSR domain $\gamma\le \gamma_{AT}$ corresponding to solutions with stable unbroken replica symmetry, $Q=0$.

\subsection{\sf Analysis of Replica Symmetry Breaking for the linear-quadratic family of encryptions.}

In this section we use the following scaling variables naturally arising when performing the analysis of the
general case of linear-quadratic family: the scaled NSR $\tilde{\gamma}=\sigma^2/J_1^2R^2$, the variables $p=t/R$ and $\tilde{Q}=Q/R^2$  and the non-linearity parameter $a=(RJ_2/J_1)^2$.

\subsubsection{Position of the de-Almeida Thouless boundary.}

Not surprisingly, the equation Eq.(\ref{stat-Q1-fin-AT}) in scaled variables simply coincides with  Eq.(\ref{p-RS-linquad}), which we repeat below for convenience of the exposition:
\begin{equation}\label{p-RS-linquadA}
(1+a) p^2\left(\tilde{\gamma}_{AT}+2(1-p)+a(1-p^2)\right)=\mu (1-p^2)(1+ap)^2\,,
\end{equation}
whereas Eq.(\ref{AT-line-gen}) takes after simple rearrangements the form
\begin{equation}\label{AT-line-linquad}
p^2=\frac{\mu}{(1+a)^3}(1+ap^2)(1+ap)^2\,.
\end{equation}
One can further use Eq.(\ref{AT-line-linquad}) to bring Eq.(\ref{p-RS-linquadA}) to
a more convenient form explicitly defining $\tilde{\gamma}_{AT}$ as:
\begin{equation}\label{gammaAT}
\tilde{\gamma}_{AT}=(1-p)^2\,\frac{(a^2+a-1)+2a(a+1)p+a^2p^2}{1+ap^2}\,.
\end{equation}
 To find $\tilde{\gamma}_{AT}$ for given values of the parameters $\mu\ge 1$ and $a\ge 0$ one has to find a value $p\in [0,1]$
by solving  Eq.(\ref{AT-line-linquad}), and substitute it to Eq.(\ref{gammaAT}). A simple consideration shows that both sides of
eq.(\ref{AT-line-linquad}), $f_L(p)=p^2$ and $f_R(p)=\frac{\mu}{(1+a)^3}(1+ap^2)(1+ap)^2$ are monotonically increasing and convex for $p\in[0,1]$, with the value of the right-hand side being larger than the left-hand side at both ends of the interval, see Fig. 3. This implies that generically there must be either no solutions if $\mu>\mu_{AT}(a)$,  or two solutions: $0\le p_{2}< p_1 < 1$ if $\mu<\mu_{AT}(a)$. The parameter $\mu_{AT}(a)$ is precisely one when only a single solution is possible, and  corresponds geometrically to the situation when the curves $f_L(p)$ and $f_R(p)$ touch each other at some $p=p_{AT}\in[0,1]$, see the Fig. 3 below.
 \begin{figure}[h!]
  \includegraphics[width=0.5\textwidth]{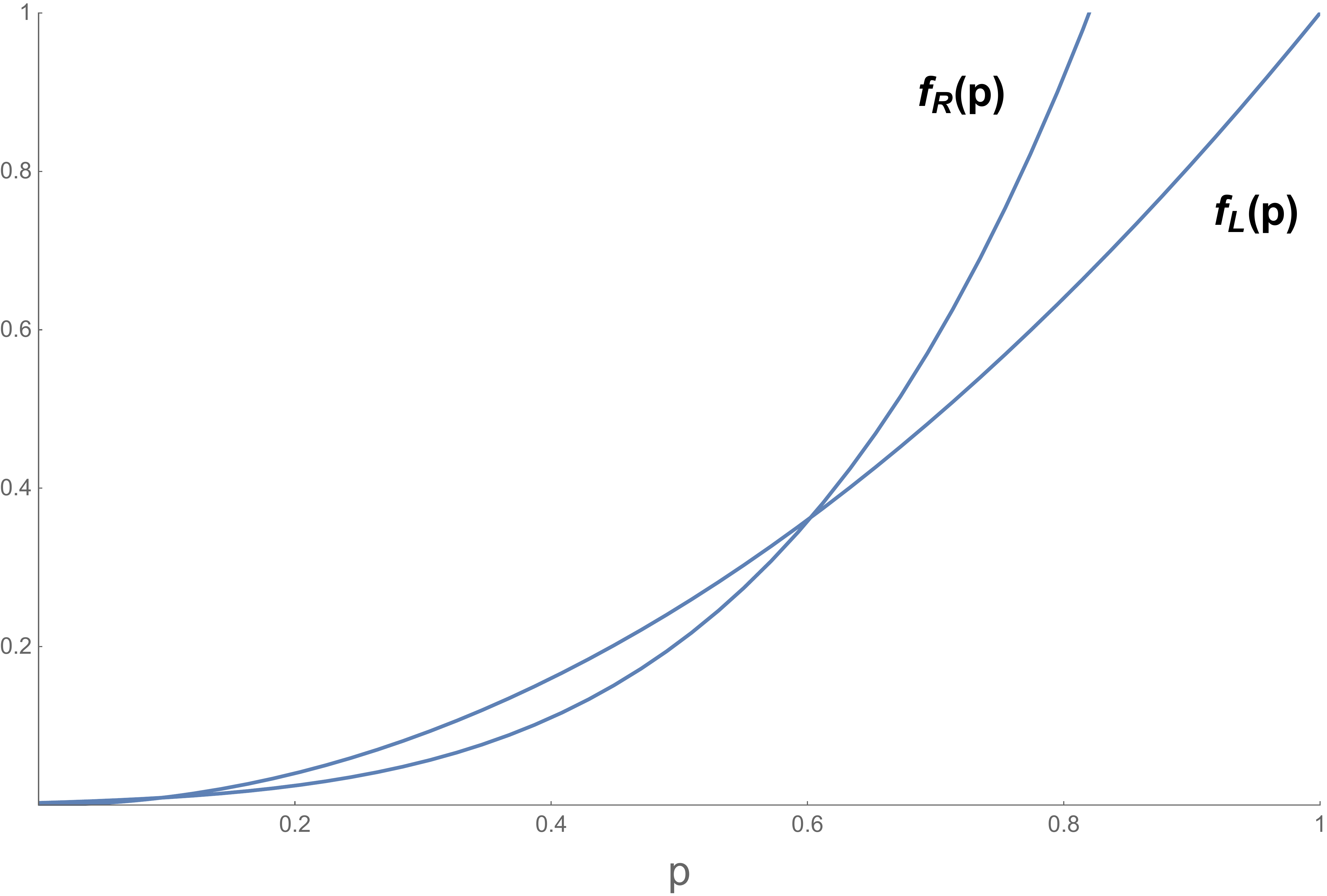} \hspace{0.5cm} \includegraphics[width=0.5\textwidth]{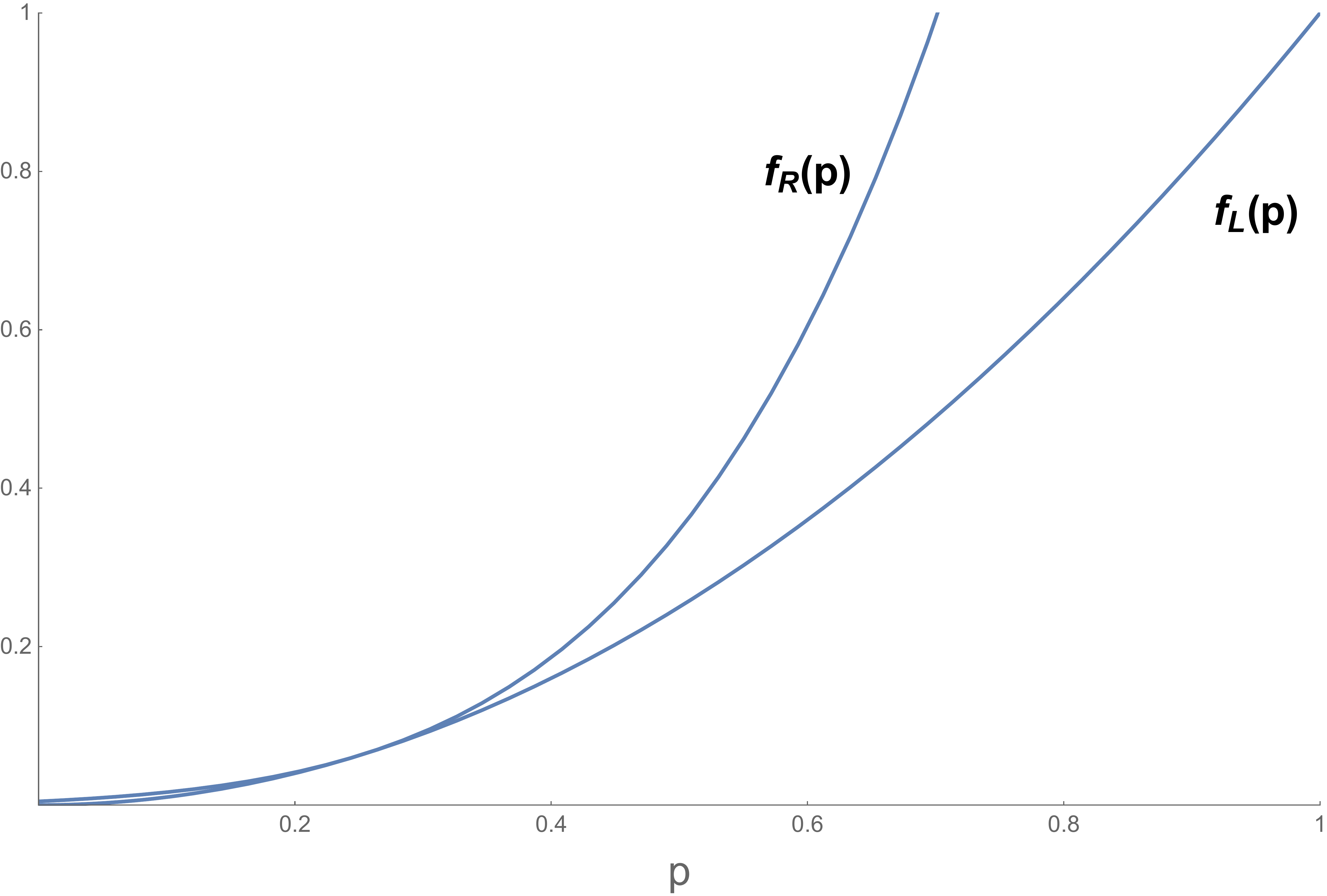}
  \caption{\label{figure1} Plots of $f_L(p)$ and $f_R(p)$ for $a=8$ and $\mu=2$ (left) and $\mu=(3/2)^3$ (right).
  In the former case the two curves intersects in two points $p_2<p_1$, in the latter case the two curves touch each other
  at $p=a^{-2/3}=0.25$. For $\mu>(3/2)^3$
  the two curves do not intersect (not shown).}
\end{figure}
The latter can be then found as  a solution to the system of two equations: $f_L(p)=f_R(p)$ and  $f'_L(p)=f'_R(p)$
for $p_{AT}$ and $\mu_{AT}$ for a given $a>1$. Surprisingly, the system can be solved explicitly:
\begin{equation}\label{muAT}
 p_{AT}=a^{-2/3}, \quad  \mu_{AT}(a)=\frac{(a^{2/3}-a^{1/3}+1)^3}{a}
\end{equation}
 A detailed mathematical analysis of the discriminant of the 4$th$-order polynomial equation Eq.({\ref{AT-line-linquad})\footnote{I am grateful to Dr. Mihail Poplavskyi for his help with the corresponding analysis.}
  fully confirms the picture outlined above, giving the explicit criterion for existence of solutions in the $(\mu,a)$ parameter plane, cf. Fig. 1:
\begin{enumerate}
\item For a given $\mu>1$ no solutions with $p\in[0,1]$ are possible for $a<1$, whereas for a fixed $a>1$ no solutions exist for $\mu>\mu_{AT}(a)$.
\item    For $\mu=\mu_{AT}(a)$ there exists a single solution: $p=p_{AT}$.
\item For a given  $a>1$ and $1<\mu<\mu_{AT}(a)$ there exist exactly two solutions $0\le p_{2}< p_1 < 1$.
\end{enumerate}
Correspondingly, in the case (i) the RS solution is valid  for all values of the scaled
Noise-to-Signal ratio $\tilde{\gamma}$, with the parameter $p$ given by solving the RS equation Eq.(\ref{p-RS-linquad}). In contrast, in the last case (iii) the two solutions give rise to two different AT thresholds in the scaled $NRS$ values: $\tilde{\gamma}^{(2)}_{AT}>\tilde{\gamma}^{(1)}_{AT}$.
In other words, for fixed values of parameters $\mu$ and $a$ there is generically an interval of NSR's $\tilde{\gamma}^{(1)}_{AT}<\tilde{\gamma}<\tilde{\gamma}^{(2)}_{AT}$ such that the replica-symmetry is broken inside and preserved for $\tilde{\gamma}$ outside that interval.

As is easy to see, for the minimal value ERP $\mu=1$ and any $a>1$ one must have only one solution at the edge of the interval: $p=1$, with  $\tilde{\gamma}_{AT}=0$. Let us increase $\mu$ slightly so that $\mu-1\ll 1$. A simple perturbation analysis then shows that a solution to Eq.(\ref{AT-line-linquad}) close to the interval edge exists, and is given by:
\be\label{smallmu}
p_1=1-\frac{a+1}{2(a-1)}(\mu-1)+o(\mu-1), \quad
\ee
We conclude that for a fixed $a>1$ and small ERP values $\mu-1\ll 1$ the replica symmetry is broken for NRS satisfying
\be\label{smallmuAT}
\tilde{\gamma}>\tilde{\gamma}^{(1)}_{AT}=\left(a-\frac{1}{4}\right)\left(\frac{a+1}{a-1}\right)^2(\mu-1)^2+o((\mu-1)^2), \quad
\ee
 Finally, one may also consider the AT equations in the limiting case of large nonlinearity $a\gg 1$ when the quadratic term in the covariance is dominant over the linear term. In this limit one easily finds two solutions of Eq.(\ref{AT-line-linquad}), given to the leading orders by $p_1^2\approx 1/\mu$ and $p_{2}\approx \frac{\sqrt{\mu}}{a^{3/2}}$ yielding the AT thresholds
\begin{equation}\label{AT_quadlim}
\tilde{\gamma}^{(1)}_{AT} \approx a\,\frac{(\mu-1)^2}{\mu}, \quad \tilde{\gamma}^{(2)}_{AT}\approx  a^2+2\sqrt{\mu\, a},
\end{equation}

We see that the ratio $\tilde{\gamma}^{(1)}_{AT}/a$ remains finite in the limit $a\to 0$, whereas  $\tilde{\gamma}^{(2)}_{AT}/a\to \infty$. To interpret this fact we recall that $a\to \infty$ is equivalent to $J_1^2\to 0$ at a fixed value of $J_2^2>0$.  Then the value $\frac{\tilde{\gamma}}{a}\equiv\frac{\gamma}{J_2^2}:=\hat{\gamma}$.
We conclude that the value $\tilde{\gamma}^{(1)}_{AT}/a=\frac{(\mu-1)^2}{\mu}:=\hat{\gamma}_{AT}$ must give the value of AT boundary in NRS for  a given ERP $\mu >1$ for the purely quadratic encryption, with the second threshold in this limiting case escaping to infinity and leaving the system in the RSB phase for all $\hat{\gamma}>\frac{(\mu-1)^2}{\mu}$. This conclusion will be fully confirmed by a detailed analysis of the purely quadratic case given in the next section.

\subsubsection{Analysis of solutions with the broken replica symmetry for the linear-quadratic family of encryptions.}
After getting some understanding of the domain of parameters where replica symmetry is expected to be broken, let us
analyse the pair of equations Eq.(\ref{stat-Q1-fin}) and Eq.(\ref{der-stat-fin}), looking for a solution with $0<Q\le 1$.
As before introduce $p=t/R$ and $\tilde{Q}=Q/R^2$  as our main variables of interest.

We start with considering the two limiting cases in the family: that of purely linear scheme with $\Phi(u)=J_1^2u$
and  the opposite limiting case of purely quadratic encryption scheme with  $\Phi(u)=J_2^2\frac{u^2}{2}$.
In the former case the previous analysis indicates that only RS solution must be possible. Indeed, we immediately notice that for purely linear scheme Eq.(\ref{der-stat-fin}) takes the form
$  p^2=\mu$ which can not have any solution as $\mu > 1$ but $p\in [0,1]$. \footnote{One can in fact easily demonstrate
that the pair  Eq.(\ref{stat-Q1-fin}) and Eq.(\ref{der-stat-fin}) can not have a real solution for \it{any} $\mu>0$.}.
We conclude that a solution with broken replica symmetry $Q>0$ does not exist, so in this case the correct value of $p$
is always given by solving the RS equation Eq.(\ref{p-RS-lin}), as anticipated.

In the opposite limiting case of purely quadratic encryption we first need to introduce a different
 scaling for NRS as $\hat{\gamma}=\frac{\gamma}{J_2^2R^2}$.  Then one may notice that
unless $p=0$ (which is always a solution) the pair  Eq.(\ref{stat-Q1-fin}) and Eq.(\ref{der-stat-fin}) reduces to the form
\begin{equation}\label{purequad-RSB}
 \fl \mu(1-p^2-\tilde{Q})=(1-\tilde{Q})\left[\hat{\gamma}+1-p^2-\tilde{Q}+\frac{1}{2}\tilde{Q}^2\right], \quad \mu p^2=(1-\tilde{Q})^3
\end{equation}
Since $p\ne 0$ implies $\tilde{Q}\ne 1$, we can further simplify this system and bring it to the form
\begin{equation}\label{purequad-RSB2}
  \tilde{Q}^3+3\left(\frac{\mu}{2}-1\right)\tilde{Q}^2+3\tilde{Q}(1-\mu)+\delta\mu=0, \quad \mu p^2=(1-\tilde{Q})^3
\end{equation}
where we introduced, in accordance with the Eq.(\ref{AT_quadlim}),
\begin{equation}\label{ATthresh}
\delta=\hat{\gamma}-\hat{\gamma}_{AT}, \quad \hat{\gamma}_{AT}:=\frac{(\mu-1)^2}{\mu}
\end{equation}
At this point we need to recall that the broken replica symmetry corresponds to $Q\in(0,R^2]$, hence $\tilde{Q}\in(0,1]$.
It is easy to show that the cubic equation in Eq.(\ref{purequad-RSB2}) may have a positive solution in that interval only for $\delta>0$.  We conclude that the replica symmetry is broken for $\hat{\gamma}>\hat{\gamma}_{AT}$ whereas for $\hat{\gamma}<\hat{\gamma}_{AT}$ the RS solution with $Q\equiv 0$ and $p$ given by  Eq.(\ref{p-RS-linquad}) remains valid.
For small $0<\delta\ll 1$ one easily finds $\tilde{Q}=\frac{\mu}{3(\mu-1})\delta+O(\delta^2)$\footnote{ The two other solutions of the cubic equation can be shown to be out of the interval $(0,1]$, see the explicit example below.}.
On the other hand one can see that the solution $\tilde{Q}(\delta)\to 1$ as $\delta\to \delta_c=\frac{3}{2}-\frac{1}{\mu}$.
so that a meaningful solution only exists in the interval $\delta\in[0,\delta_c]$. Moreover, it is easy to show that
for $\delta\to \delta_c$ we have $\tilde{Q}=1-\sqrt{\frac{2}{3}(\delta_c-\delta)}$. We see that the second of Eq.(\ref{purequad-RSB2}) then implies that when approaching the true threshold value $\tilde{\gamma}^{(RSB)}_{c}=\mu-\frac{1}{2}$ dictated by broken replica symmetry the quality parameter vanishes as $p\sim (\hat{\gamma}^{(RSB)}_{c}-\hat{\gamma})^{3/4}$ rather than as a square root, as in the replica-symmetric solution Eq.(\ref{p-RS-quad}).

To study the behaviour of the solution $\tilde{Q}(\delta)$ for $\delta$ of the order of one it is instructive to
consider a particular (but generic) case of $\mu=2$, when $\delta_c=1$. The cubic equation for $\tilde{Q}$ takes then a particular simple form:
\begin{equation}\label{examp}
\tilde{Q}^3-3\tilde{Q}+2\delta=0
\end{equation}
which represents one of the rare instances when the Cardano formula for solving cubic equations is really helpful for the analysis.
Indeed, according to the Cardano formula in this case the solution is given by
 \begin{equation}\label{Cardano}
\tilde{Q}=\left[-\delta+\sqrt{\delta^2-1}\right]^{1/3}+\left[-\delta-\sqrt{\delta^2-1}\right]^{1/3}
\end{equation}
As $\delta\in [0,1]$ we can further parametrize $\delta=\sin{\phi}, \, \phi\in [0,\pi/2]$, and obtain the three different solutions to Eq.(\ref{examp}) in the following form
\begin{equation}\label{Cardano1}
\tilde{Q}_n=2\cos{\left(\frac{1}{3}\left[\phi+\pi\left(2n+\frac{1}{2}\right)\right]\right)}, \quad n=0,1,2
\end{equation}
We see then that $\tilde{Q}_{0,1}$ are outside the interval $[0,1]$, as
$\tilde{Q}_0=2\cos{\left(\frac{1}{3}\phi+\frac{\pi}{6}\right)}\in [1,\sqrt{3}]$ and
$\tilde{Q}_1=2\cos{\left(\frac{1}{3}\phi+\frac{\pi}{6}\right)}\in [-\sqrt{3},-1]$, whereas
$\tilde{Q}_2=2\sin{\frac{\phi}{3}}$ is exactly the valid solution.

The nature of the solution for purely quadratic scheme for a general $\mu>1$ is exactly of the same type.
 After finding $\tilde{Q}$ from the cubic equation for $\tilde{\gamma}_{AT}<\tilde{\gamma}\le \tilde{\gamma}^{(RSB)}_c=\mu-1/2$ we find the quality
parameter $p$ from the second of Eq.(\ref{purequad-RSB2}), and combining it with RS expression Eq.(\ref{p-RS-quad}) obtain
the full corresponding curve for $p(\hat{\gamma})$ for a given $\mu$. In particular, for the above special value $\mu=2$
the full curve can be described by an explicit expression:
\begin{equation}\label{p-Full-quad}
 p=\left\{\begin{array}{cc} \sqrt{1-\hat{\gamma}} &  \mbox{ if}\,\, \hat{\gamma}\le \hat{\gamma}_{AT}=\frac{1}{2}\\
 \frac{1}{\sqrt{2}}\left[1-2\sin{\frac{1}{3}\left(\arcsin{\left(\hat{\gamma}-\frac{1}{2}\right)}\right)}\right]^{3/2} &  \mbox{ if}\,\,
 \frac{1}{2}\le  \hat{\gamma}\le \hat{\gamma}^{(RSB)}_{c}=\frac{3}{2}\\
 0 &  \mbox{ if}\,\,\hat{\gamma} >\hat{\gamma}^{(RSB)}_c=\frac{3}{2}
 \end{array} \right.
\end{equation}
and is depicted in the left figure below. For $\mu\ne 2$ analytic solution of the cubic equation is less instructive,
and it is easier to solve the equation numerically.
 \begin{figure}[h!]
  \includegraphics[width=0.5\textwidth]{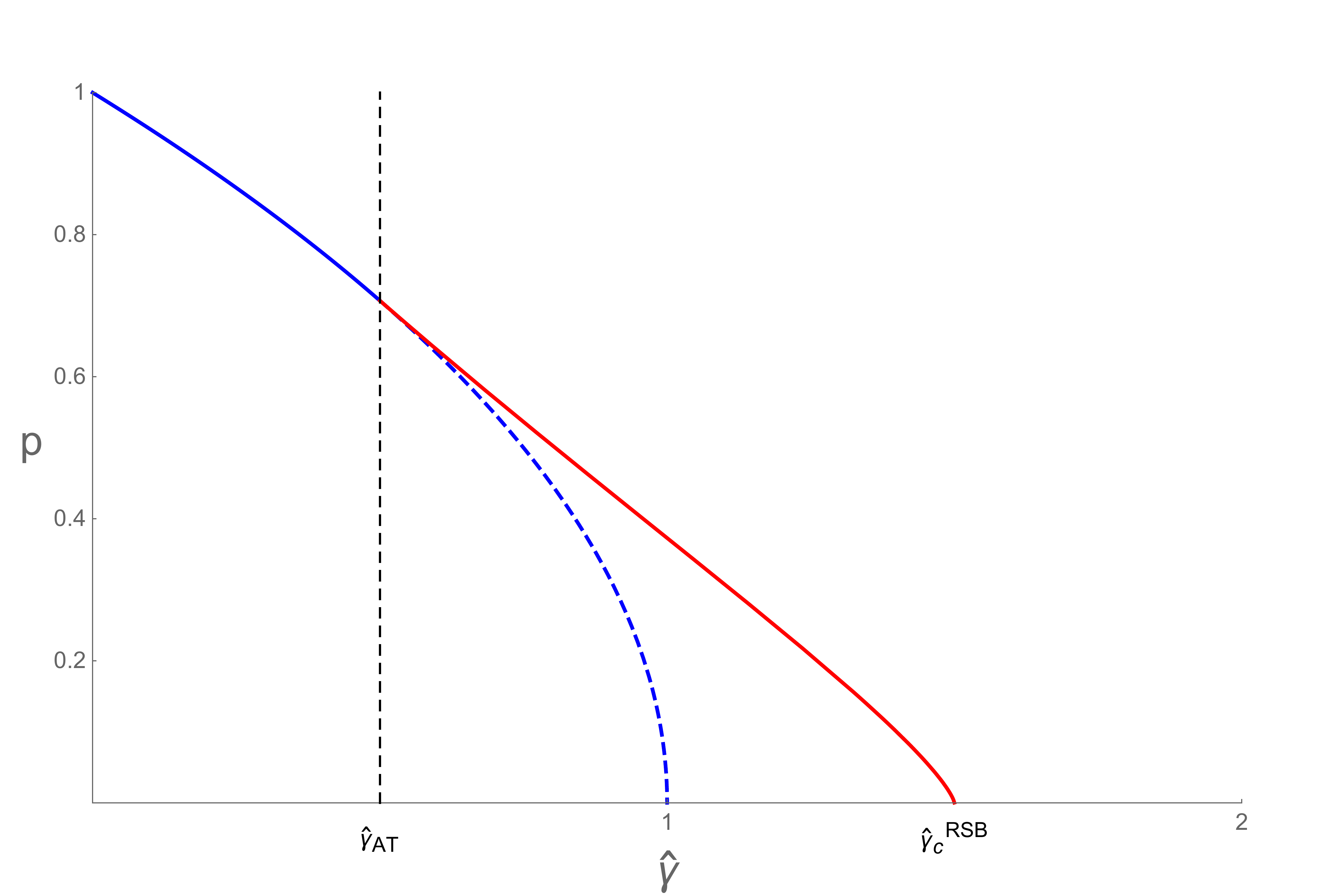}\hspace{0.5cm}\includegraphics[width=0.5\textwidth]{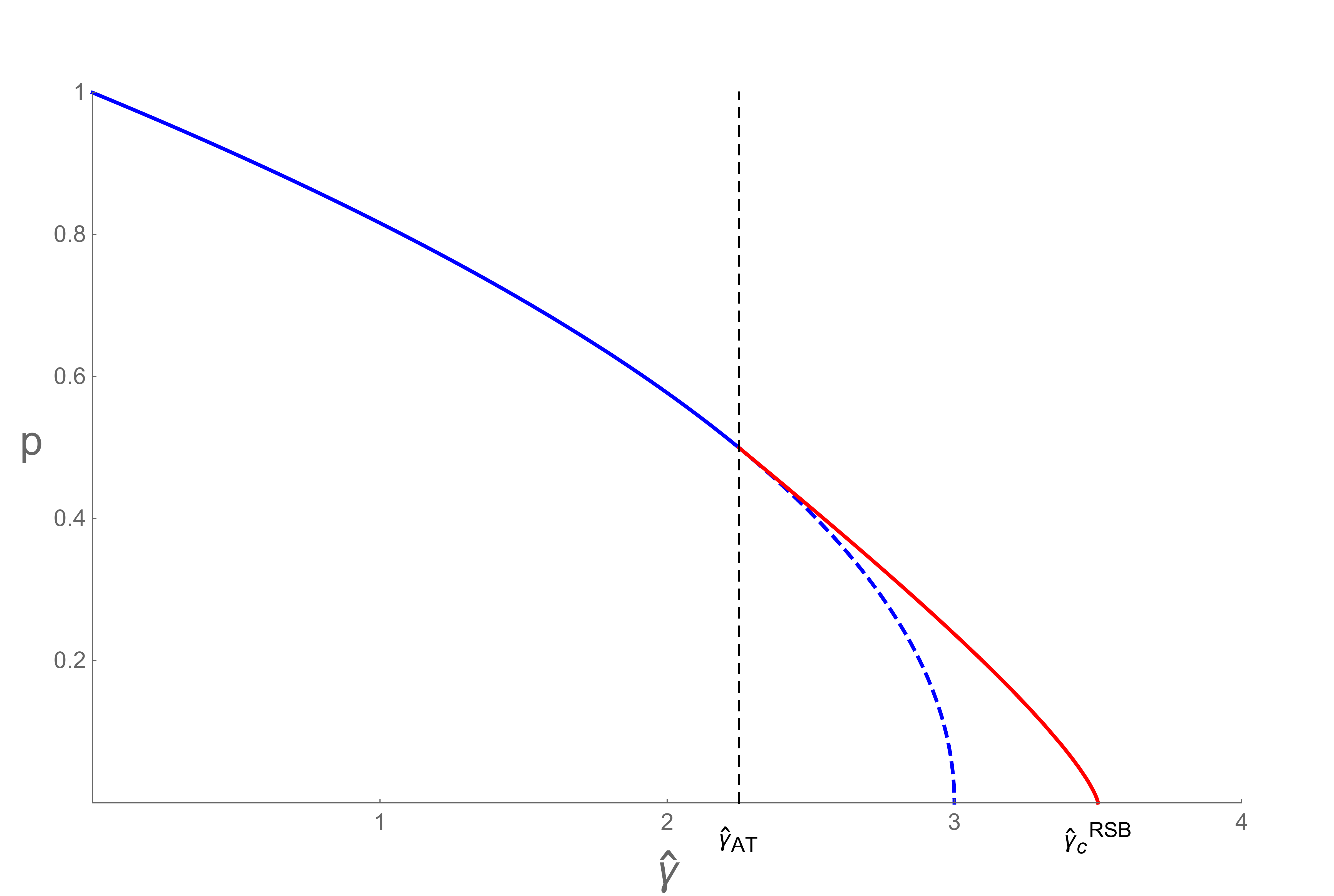}
  \caption{\label{figure1} The quality parameter $p$ as a function of the scaled noise-to-signal ratio $\hat{\gamma}$ for purely quadratic encryptions and two different values of the Encryption Redundancy Parameter, $\mu=2$ (left) and $\mu=4$ (right).
  The blue broken curve is the continuation of the replica-symmetric solution in the region of Full RSB.}
\end{figure}

After full understanding of the limiting cases we briefly discuss the solution
for a generic linear-quadratic encryption algorithm with
some finite value $1<a<\infty$ of the nonlinearity parameter.  In this case the NRS scaling
is still given by the NRS variable $\tilde{\gamma}=\gamma/J_1^2$. One recalls that for a given $a$ as long as
$\mu<\mu_{AT}(a)=\frac{(a^{2/3}-a^{1/3}+1)^3}{a}$ there exists
two NSR thresholds $\tilde{\gamma}_{AT}^{(1)}$ and $\tilde{\gamma}_{AT}^{(2)}$ such that
for $\tilde{\gamma}\notin [\tilde{\gamma}_{AT}^{(1)}, \tilde{\gamma}_{AT}^{(2)}]$
the curve $p(\tilde{\gamma})$ is given by RS solution Eq.(\ref{p-RS-linquad}),
whereas for $\tilde{\gamma}\in [\tilde{\gamma}_{AT}^{(1)}, \tilde{\gamma}_{AT}^{(2)}]$
the curve $p(\tilde{\gamma})$ is given by the Full RSB solution from the system of two equations:
\begin{equation}\label{FRSB-linquad-1}
p^2=\frac{\mu}{\left[1+a(1-\tilde{Q})\right]^3}(1+ap^2)(1+ap)^2
\end{equation}
and
\begin{equation}\label{FRSB-linquad-2}
\fl \left[1+a(1-\tilde{Q})\right] p^2\left(\tilde{\gamma}+2(1-p)+a(1-p^2)-a\tilde{Q}+a\frac{1}{2}\tilde{Q}^2\right)=\mu (1-p^2-\tilde{Q})(1+ap)^2
\end{equation}
In Fig. 5 we plot the full resulting curve.

\begin{figure}[h!]
\centering
  \includegraphics[width=0.6\textwidth]{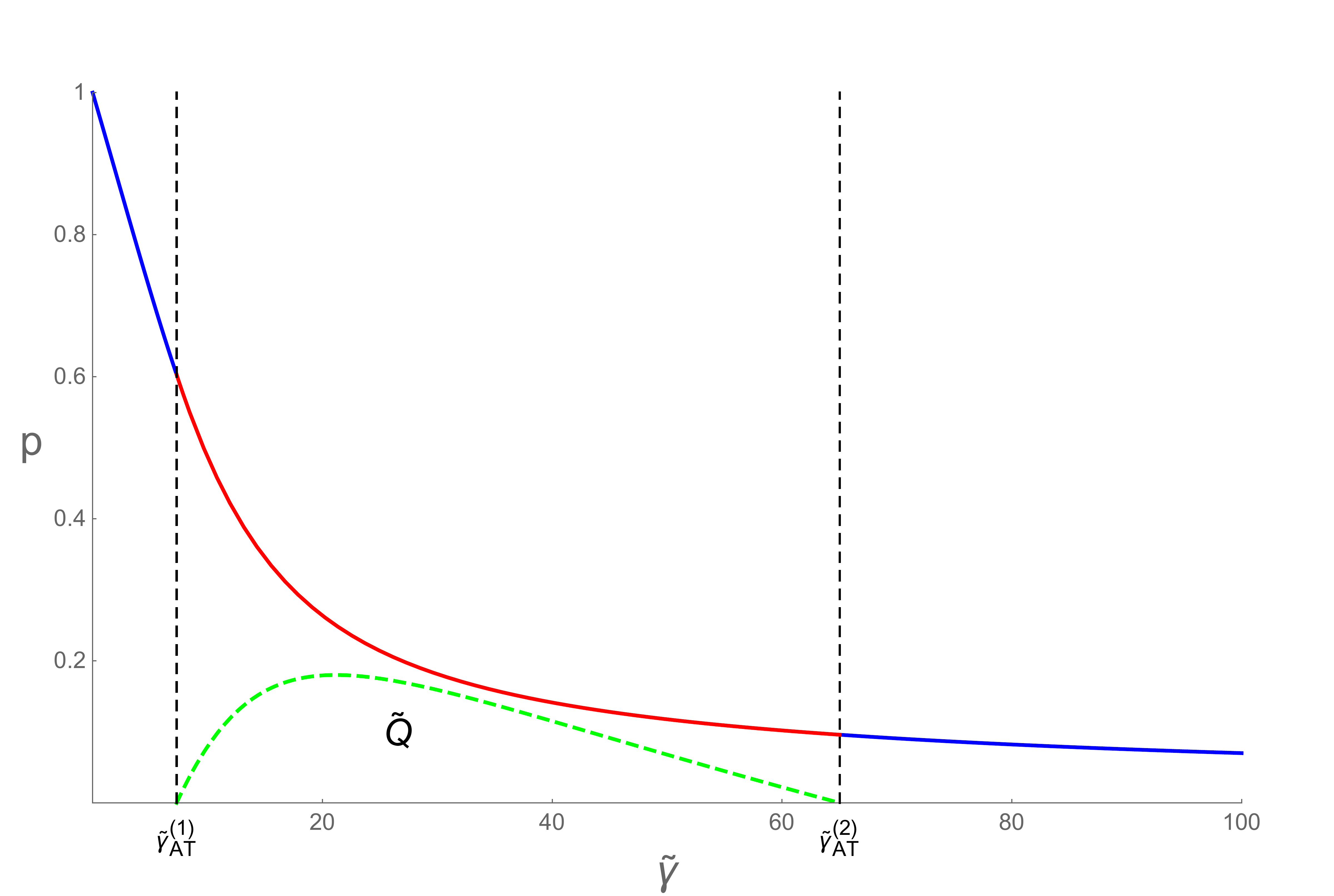}
  \caption{\label{figure1} The quality parameter $p$ as a function of the scaled noise-to-signal ratio $\tilde{\gamma}$ for a generic representative of linear-quadratic encryptions for the nonlinearity $a=8$ and redundancy $\mu=2$. In the interval of scaled noise-to-signal ratio $\tilde{\gamma}^{(AT)}_2<\tilde{\gamma}< \tilde{\gamma}^{(AT)}_2$ the replica symmetry is broken as signified by a non-zero values of the parameter $\tilde{Q}$, plotted as a green broken line.}
\end{figure}

\section{Appendix A: Perturbation theory in the Lagrange multipliers framework. }
We set the parameter  $J_1^2=1$ in this Appendix.

Substituting the solution Eq.(\ref{Lagrangesol}) into the spherical constraint $({\bf x},{\bf x})=NR^2$ and using
the notations $W:=A^TA$ and $G=\left(W-\lambda {\bf 1}_N\right)^{-1}$ one gets an equation for the Lagrange multiplier $\lambda$:
\begin{equation}\label{LagrangesolA}
\fl 2\lambda \left({\bf s}, G {\bf s}\right)+2\left({\bf s}, G\, A^T{\bf b}\right)+\lambda^2\left({\bf s}, G^2 {\bf s}\right)
+2\lambda\left({\bf s}, G^2\, A^T{\bf b}\right)+\left({\bf b},\, A \, G^2\, A^T{\bf b}\right)=0
\end{equation}
Recall that for $\sigma=0$ the global minimum corrersponds to $\lambda=0$, so we will look for a weak-noise expansion
$\lambda=\lambda_1\sigma+\lambda_2\sigma ^2+\ldots\ll 1$. Writing ${\bf b}=\sigma{\bf \xi}$, with components of ${\bf \xi}$ having variance unity, and remembering that for $M>N$ the matrix $W>0$ with probability tending to one is invertable,
 one can safely expand $G\approx W^{-1}+\lambda\,W^{-2}+ ...$ and substituting this to Eq.(\ref{LagrangesolA}) find the first and then second order coefficient in the Lagrange multiplier as:
 \begin{equation}\label{lambdacoeff11}
 \fl \lambda_1=-\frac{\left({\bf s}, W^{-1}\, A^T{\bf \xi}\right)}{p_1},\quad \lambda_2=-\frac{1}{2p_1}
 \left[3\lambda_1^2p_2+4\lambda_1\left({\bf s},W^{-2} A^T{\bf \xi}\right)+
 \left({\bf \xi},AW^{-2}A{\bf \xi}\right)\right]
  \end{equation}
 where we introduced the following notations:
 \begin{equation}\label{p1p2}
 p_1=\left({\bf s}, W^{-1}\,{\bf s}\right), \quad p_2=\left({\bf s}, W^{-2}\,{\bf s}\right)
 \end{equation}
 Using this one can get the following expansion for the quality parameter Eq.(\ref{qualirec}):
  \begin{equation} \label{qualitypert1}
\fl 1-p_N=\frac{1}{NR^2}\left\{\sigma\left[\lambda_1p_1+\left({\bf s}, W^{-1}\, A^T{\bf \xi}\right)\right]+\sigma^2\left[\lambda_1^2p_2+\lambda_2p_1+\lambda_1\left({\bf s}, W^{-2}\, A^T{\bf \xi}\right)\right]+\ldots\right\}
 \end{equation}
 valid at every realization of both the noise and the random matrix $A$. Substituting here Eq.(\ref{lambdacoeff11}) and taking the
 expected value first only over the Gaussian noise ${\bf \xi}$ gives after straightforward, but somewhat lengthy manipulations, to the leading order:
\begin{equation}\label{qualitypert2}
 1-p_N=\frac{\sigma^2}{NR^2}p_1\left\langle\lambda_2\right\rangle_{{\bf \xi}}=\frac{\sigma^2}{2NR^2}\left[Tr\left[ W^{-1}\right]-\frac{p_2}{p_1}\right]
\end{equation}
It remains to perform the average over the ensemble of Wishart matrices $W$ with $M>N$. One finds that as long as $N\gg 1$ the fraction $p_2/p_1$ remains of the order of unity, whereas the first term  $Tr\left[ W^{-1}\right]=O(N)$ and is dominant.
Using the well-known Marchenko-Pastur limiting law for the spectral density of eigenvalues of $W$
\begin{equation}\label{MP}
\rho_{MP}(\lambda)=\frac{2}{\pi}\frac{1}{\left(\sqrt{\lambda_{+}}-
\sqrt{\lambda_{-}}\right)^2}\,\frac{\sqrt{(\lambda-\lambda_{-})(\lambda_{+}-\lambda)}}{\lambda}, \quad \lambda_{-}\le \lambda\le \lambda_{+}
\end{equation}
where $\lambda_{\pm}=(\sqrt{\mu}\pm 1)^2$ are positions of the spectral edges,  one can find for the mean trace of the resolvent:
\begin{equation}\label{WishartRes}
\fl \lim_{N\to \infty}\frac{1}{N}\left\langle\left[(W-\lambda{\bf 1}_N)^{-1}\right]\right\rangle=
\frac{2}{\sqrt{\lambda_{-}\lambda_{+}}-\lambda+\sqrt{(\lambda-\lambda_{-})(\lambda_{+}-\lambda)}}, \quad \lambda<\lambda_{-}
\end{equation}
In particular, for $\lambda=0$ we have
\begin{equation}\label{WishartResZERO}
 \lim_{N\to \infty}\frac{1}{N}\left\langle  Tr\left[ W^{-1}\right]\right\rangle= \frac{1}{\sqrt{\lambda_{-}\lambda_{+}}}=\frac{1}{\mu-1}
\end{equation}
resulting in Eq.(\ref{qualiasylin}) valid in the first-order in small-noise value.\\

One can also straightforwardly extract the behaviour for asymptotoically large noise variance values $({\bf b},{\bf b})\propto \sigma^2 \to \infty$ when Eq.(\ref{Lagrangesol})
implies that
 \begin{equation}\label{largenoise}
 {\bf x  }\approx \left(W-\lambda {\bf 1}_N\right)^{-1}\,A^T{\bf b}
\end{equation}
 Substituting Eq.(\ref{largenoise}) to the spherical constraint gives
 \begin{equation}\label{largenoise1}
 ({\bf x},{\bf x})={\bf b}^T\, A\left(W-\lambda {\bf 1}_N\right)^{-2}\,A^T{\bf b}=NR^2
 \end{equation}
 and further averaging over the Gaussian noise ${\bf b}$ the above relation yields:
 \begin{equation}\label{largenoise2}
 1=\frac{\sigma^2}{NR^2}\, Tr\left[A\left(W-\lambda {\bf 1}_N\right)^{-2} A^T\right]=\frac{\sigma^2}{NR^2}\, Tr\left[\left(W-\lambda {\bf 1}_N\right)^{-2}W\right]
 \end{equation}
 It is clear that for $\tilde{\gamma}=\frac{\sigma^2}{R^2}\gg 1$  the relevant value of the Lagrange multiplier $\lambda$ has to be large in modulus, $|\lambda|\gg 1$, which immediately implies in the large-$N$ limit:
 \begin{equation}\label{largenoise3}
\fl \lambda^2 \approx\tilde{\gamma}\, {\small \frac{1}{N}} Tr\,W\,|_{N\to \infty} =\tilde{\gamma} \int_{\lambda_{-}}^{\lambda_{+}}\rho_{MP}(\lambda)\,\lambda\, d\lambda= \tilde{\gamma}\frac{(\sqrt{\lambda_+}+\sqrt{\lambda_-})^2}{4}=\tilde{\gamma}\mu
 \end{equation}
 so that $|\lambda| \approx \sqrt{\tilde{\gamma}\mu}\gg 1$.  Now, the Eq.(\ref{Lagrangesol}) implies for the quality parameter
 \begin{equation}\label{qual_largenoise}
\fl p_N=\frac{({\bf x},{\bf s})}{NR^2}=1-\frac{\lambda}{NR^2} {\bf s}^T  \left(W-\lambda {\bf 1}_N\right)^{-1}{\bf s}+ \frac{1}{NR^2} {\bf s}^T  \left(W-\lambda {\bf 1}_N\right)^{-1}A^T {\bf b}
 \end{equation}
 Expanding for large $|\lambda|\sim \sqrt{\gamma}\gg 1$  as
 \[
\fl {\bf s}^T  \left(W-\lambda {\bf 1}_N\right)^{-1}{\bf s}\approx -\frac{NR^2}{\lambda}+\frac{1}{\lambda^2} \left({\bf s}, W{\bf s}\right)+\ldots, \quad {\bf s}^T  \left(W-\lambda {\bf 1}_N\right)^{-1}A^T {\bf b}\approx -\frac{1}{\lambda}{\bf s}^T  A^T {\bf b}+\ldots
 \]
shows that to the leading order
 \begin{equation}\label{qual_largenoise1}
 p_N=-\frac{1}{\lambda}\frac{1}{NR^2}\left[\left({\bf s}, W{\bf s}\right)+\left({\bf s}, A^T {\bf b}\right)\right]
 \end{equation}
 which upon averaging over the Wishart matrices $W$ and the noise ${\bf b}$, taking the limit $N\to \infty$,  and taking into account that actually $\lambda\to -\infty$, yields
 \begin{equation}\label{largenoise-lead-asy}
 p_{\infty}=\frac{1}{|\lambda|}\frac{1}{N} \left\langle Tr\,W\right\rangle \,|_{N\to \infty}=\frac{\mu}{\sqrt{\mu\tilde{\gamma}}}=\sqrt{\frac{\mu}{\tilde{\gamma}}}
 \end{equation}

 \section{Appendix B}
In this Appendix we give a proof of the following

{\bf Lemma}

{\it Let ${\bf x}\in \mathbb{R}^N$, and $V({\bf x})$ be a Gaussian random field with mean zero and the covariance
  \begin{equation}\label{covgen}
\left\langle V({\bf x}_1)V({\bf x}_2)\right\rangle=\phi\left({\bf x}_1;{\bf x}_2\right)
\end{equation}
where $\phi({\bf x};{\bf y})=\phi({\bf y};{\bf x})$ is any suitable covariance structure function. Then for any $\beta>0$ holds
\be\label{multichigen}
\left\langle e^{-\frac{\beta}{2} \sum_{a=1}^n V^2({\bf x}_a)} \right\rangle
=\left[\det{{\cal G}({\bf x}_1,\ldots,{\bf x}_n}\right]^{-1/2}
\ee
where the (positive definite) $n\times n$ matrix ${\cal G}({\bf x}_1,\ldots,{\bf x}_n)$ has the entries
 \be\label{Ggen}
 {\cal G}_{ab}({\bf x}_1,\ldots,{\bf x}_n)=\delta_{ab}+\beta\phi({\bf x}_a;{\bf x}_b)
\ee
}
{\bf Proof}: it is convenient  to linearize the squared  terms in the exponential by exploiting the Gaussian integration of an auxiliary real variable $u_a$ for every $a=1,\ldots, n$ (the trick known in the physical literature as the Hubbard-Stratonovich transformation):
\begin{equation}\label{HS}
e^{-\frac{\beta}{2}V^2({\bf x}_a)}=\int_{\mathbb{R}}e^{-i\sqrt{\beta}u_a
V({\bf x}_a)}\,\,\frac{e^{-\frac{1}{2}u_a^2}\,du_a\,}{\sqrt{2\pi}}
\end{equation}
which implies
 \be\label{HS1}
 \left\langle e^{-\frac{\beta}{2} \sum_{a=1}^n V^2({\bf x}_a)} \right\rangle
=\int_{\mathbb{R}^n}\left\langle e^{-i\sqrt{\beta}\sum_{a=1}^n u_a
V({\bf x}_a)}\right\rangle\,\,e^{-\frac{1}{2}\sum_{a}u_a^2}\,\prod_a \frac{du_a}{\sqrt{2\pi}}
\ee
Now the average is immediate to perform due to the Gaussian nature of the random field $V({\bf x})$.
Using Eq.(\ref{covgen}) we get:
\begin{equation}\label{disave}
 \left\langle e^{-i\sqrt{\beta} \sum_{a=1}^n\, u_a V({\bf x}_a)}\right\rangle=
e^{-\frac{\beta}{2}\sum_{a,b}^n u_a u_b\phi({\bf x}_a,{\bf x}_b)}
\end{equation}
Substituting now Eq.(\ref{disave}) back to Eq.(\ref{HS1}) we see that the integrals over
the variables $u_a$ remain multivariate Gaussian, and hence can be easily performed, resulting in Eq.(\ref{multichigen}).

\section{APPENDIX C}

In this Appendix we give a proof of the following

{\bf Theorem}

{\it
Consider a function $ F({\bf x_1},...,{\bf x_n};{\bf s})$
of $N$-component real vectors ${\bf x}_l\,\, 1\le l\le n$ and a $N$-component real vector ${\bf s}$ (considered as a parameter)
such that
\begin{equation}\label{conv1}
\int_{\mathbb{R}^N}d{\bf x}_1...\int_{\mathbb{R}^N}d{\bf x}_n
|F({\bf x_1},...,{\bf x_n};{\bf s})|<\infty
\end{equation}
Suppose further that the function $F$
depends on its arguments only via $n(n+1)/2$ scalar products $\tilde{Q}_{ab}=({\bf x}_a,
{\bf x}_{b}), \, 1\le a,b\le n$ and on $n$ projections $\tilde{t}_a=({\bf x}_a,
{\bf s})$ for $a=1,\ldots, n$.    Rewrite then such a function as
${\cal F}(\tilde{Q}, \tilde{{\bf t}})$ of
$n\times n$ real symmetric matrix $\tilde{Q}$ with entries $\tilde{Q}_{ab}$ and a vector ${\bf t}=(t_1,\ldots, t_n)$.
Then for $N\ge n+1$ the integral defined as
\begin{equation}\label{defint}
I^{(F)}_{N,n}({\bf s})=\int_{\mathbb{R}^N}d{\bf x}_1... \int_{\mathbb{R}^N}d{\bf x}_n
F({\bf x}_1,...,{\bf x}_n;{\bf s})\,
\end{equation}
is equal to
\begin{equation}\label{trans1}
I^{({\cal F})}_{N,n}({\bf s})={\cal C}^{(o)}_{N-1,n}
\int_{Q>0}
{\cal F}(Q+{\bf t}\otimes {\bf t}^{T}, |{\bf s}|{\bf t})\,\left[\det{Q}\right]^{(N-n-2)/2}\,dQ\,d{\bf t}
\end{equation}
where the proportionality constant is given by
\be \label{const}
{\cal C}^{(o)}_{N,n} = \frac{\pi^{\frac{n}{2}\left(N-\frac{n-1}{2}\right)}}
{\prod_{k=0}^{n-1}\Gamma\left(\frac{N-k}{2}\right)},\,
\ee
 the integration in the Eq.(\ref{trans1})
goes over the manifold
of real symmetric non-negative definite $n\times n$ matrices $Q$ and the vector ${\bf t}\in \mathbb{R}^n$,
whereas the diadic product $T={\bf t}\otimes {\bf t}^{T}$ is used to denote a (rank one) $n\times n$ matrix $T$ with entries
$T_{ab}=t_at_b$.
}

{\bf Proof}: Denote ${\bf e}_N=(0,\ldots,0,1)$ the last of the standard basis vectors in $\mathbb{R}^N$. Then there exists an orthogonal transformation $O({\bf s})\in O(N)$ such that we can
represent the vector ${\bf s}$ as ${\bf s}=|{\bf s}|O({\bf s}) {\bf e}_N$. Perform the transformation of variables ${\bf x}_a\to
{\bf y}_a=O({\bf s}) {\bf x}_a,\,\, \forall a=1,\ldots,n$ in the integrand of Eq.(\ref{defint}).
Such transformation leaves invariant the volume element: $\prod_a d{\bf x}_a=\prod_a d{\bf y}_a$ and  the scalar products $\tilde{Q}_{ab}=({\bf x}_a,
{\bf x}_{b})=({\bf y}_a,
{\bf y}_{b}), \, 1\le a,b\le n$ but transforms the $n$ projections $\tilde{t}_a=({\bf x}_a,
{\bf s})$ for all $a=1,\ldots, n$ into $\tilde{t}_a=|{\bf s}|y_{aN}$, where $y_{aN}$ is the $N-$th component of the vector ${\bf y}_a$. Now decompose each vector ${\bf y}_a$ as ${\bf y}_a=({\bf v}_a,y_{aN})$, where ${\bf v}_a$ are $(N-1)$-dimensional vectors.
Such a decomposition implies:
\[
\tilde{Q}_{ab}=({\bf y}_a,
{\bf y}_{b})=({\bf v}_a,
{\bf v}_{b})+y_{aN}y_{bN}, \quad \prod_a d{\bf y}_a=\prod_a d{\bf v}_a\, \prod_a d y_{Na}
\]
 so that using the notations of the Theorem, renaming $y_{Na}\to t_a$ and introducing ${Q}_{ab}^{({\bf v})}=({\bf v}_a,
{\bf v}_{b})$ and ${\bf t}=(t_1,\ldots,t_n)$ we can rewrite Eq.(\ref{defint}) as
 \begin{equation}\label{intermed}
\fl I^{(F)}_{N,n}({\bf s})=\int_{\mathbb{R}^n} \mathcal{I}^{(F)}_{N,n}({\bf t};{\bf s})\,d{\bf t}, \quad
\mathcal{I}^{(F)}_{N,n}({\bf t};{\bf s}):=\int_{\mathbb{R}^{N-1}}d{\bf v}_1... \int_{\mathbb{R}^{N-1}}d{\bf v}_n
{\cal F}\left(Q^{({\bf v})}+{\bf t}\otimes {\bf t}^{T}, |{\bf s}|{\bf t}\right)
\end{equation}
Note that the last integral with respect to vectors ${\bf v}_a,\, a=1,\ldots, n$ has the full $O(N-1)$ invariance of the integrand.
The statement of the Theorem then immediately follows by applying to this situation the 'dimensional reduction' formula suggested  for the first time in \cite{Percus87} and essentially rediscovered in \cite{F2002a};  see the Appendix D of \cite{F2002b} and the appendix B of \cite{logmultlec} for alternative proofs.

\end{document}